\documentclass[11pt]{article}

\usepackage{amsmath,amssymb,amsthm}
\usepackage{braket}
\usepackage{booktabs}
\usepackage{multirow}
\usepackage[table]{xcolor}
\usepackage{geometry}
\usepackage{hyperref}
\usepackage{float}
\usepackage{graphicx}
\usepackage{subcaption}
\newcommand{\ii}{\mathrm{i}}

\begin{document}


\begin{center}
{\LARGE \bfseries Knot your average qutrit: Measurement-induced entanglement splitting and the cabling dictionary for $GHZ$ and $W$ States}

\vspace{1.2em}

{\large Sougata Bhattacharyya$^{1}$}

\textit{$^{1}$Department of Astronomy, Astrophysics and Space Engineering,}\\
\textit{Indian Institute of Technology, Indore, 453552, Madhya Pradesh, India}

\vspace{0.8em}

{\large Sovik Roy$^{2}$}

\textit{$^{2}$Department of Mathematics, Techno Main Salt Lake,}\\
\textit{Techno India Group, EM 4/1, Sector V, Salt Lake, Kolkata 700091, India}
\end{center}

\vspace{0.5em}

\vspace{0.5em}
\hrule
\vspace{1.5em}

\begin{abstract}
\noindent Multipartite entanglement is conventionally classified by state families viz. family of $GHZ$ and $W$ class of states, with each family expected to behave differently under measurement. We show that, at least for the question of how entanglement splits after a single-particle measurement, this is not the division that matters for qutrits. Extending the Aravind's correspondence (which models entanglement as topological linking, and projective measurement as physically cutting a ring from an interlinked configuration\cite{aravind1997}) from qubits to qutrits, we derive the complete measurement-induced entanglement splitting of the $GHZ$ type qutrit state i.e $\ket{GHZ_3}$  and of the full family of symmetric $W$ class qutrit states, six \textit{two same - one different} states i.e $\ket{W_{p,p,q}^{sym}}$ and one \textit{all - different} state i.e. $\ket{W_{\{0,1,2\}}}$, under both the computational basis (CB) and the mutually unbiased bases (MUBs), obtaining exact eigenvalues and Schmidt ranks for every outcome in every case. We see that the $\ket{W_{\{0,1,2\}}}$ state behaves similarly as $\ket{GHZ_3}$ state, a single, outcome-independent residual rank in each basis, while the $\ket{W_{p,p,q}^{sym}}$ states alone show
probability-weighted, outcome-dependent behaviour. The relevant structural line is therefore \textit{repeated-index} versus \textit{all-different-index} bag structure, not $GHZ$ class versus $W$ class. We express this classification using a \textit{two-strand cabling} extension of Aravind's \textit{ring-and-link} picture. This is needed because the qutrit residual Schmidt rank ($R$) takes three values, $R \in \{1~,~2,~3\}$, rather than the qubit binary (i.e. $R \in \{1~,~2\}$). We are explicit throughout that \textit{this cabling dictionary} is a labeling convention built to reproduce an independently computed Schmidt rank, not a topological invariant derived from the link diagrams themselves, and we discuss what would be needed to close that gap.

\end{abstract}

\vspace{1em}

\vspace{1em}
\hrule
\vspace{1em}
\newpage
\tableofcontents
\newpage


\section{Introduction}
\label{sec:introduction}

\noindent Quantum entanglement is conventionally characterized through algebraic invariants such as the Schmidt rank \cite{eisert2001,sperling2011} and concurrence \cite{wootters1998,coffman2000}, that are extracted directly from a state's wavefunction. The properties, measures, and characterization of such invariants across bipartite and multipartite systems having been extensively reviewed \cite{horodecki2009quantum, amico2008, horodecki2001entanglement, plenio2014introduction}. A complementary, highly visual framework instead treats multipartite entanglement as a form of topological linking. Pioneered by Aravind for multi-qubit systems \cite{aravind1997} and developed further by Kauffman and Lomonaco \cite{kauffman2002quantum}, this correspondence maps each particle to a closed physical ring, modeling projective measurement as the physical severing and removal of that ring from the overall configuration. Related topological classifications of entanglement have since been proposed independently by other groups, extending the correspondence to broader multiqubit settings \cite{li2008entanglement, quinta2018classifying}. This geometric perspective beautifully captures the all-or-nothing correlations of the three-qubit Greenberger–Horne–Zeilinger ($GHZ$) state \cite{GHZ1989}, which maps exactly to the Borromean rings \cite{aravind1997} severing any one ring completely unlinks the remaining two. However, extending this rigid structural analogy to the three-qubit $W$ state \cite{Dur2000} introduces immediate complications. Because a measured $W$ state retains bipartite entanglement with a probability dictated by the measurement outcome, its topological counterpart cannot be a fixed, static invariant, but rather a probability-weighted configuration that demands a more flexible descriptive vocabulary \cite{kauffman2019topological}.\\

\noindent Our previous work demonstrated that the resilience of multipartite entanglement under local measurements admits a natural correspondence with topological link structures such as the 3-Hopf link, 3-link chain, and Borromean rings\cite{sougata2509}. In another work, we developed this topological dictionary further for symmetric Dicke states of qubits, introducing the notion of \emph{link fluidity} to describe how a fixed link type can persist, weaken, or reorganize under different single-particle measurements \cite{sougata2512}. Nevertheless, we established a measurement surgery correspondence by providing a geometric topological framework for single-qubit Pauli measurements on one-dimensional linear cluster states, with framed ribbon links capturing both connectivity and quantum phase \cite{sougata2602}. These analyses remained entirely within the qubit setting. The residual Schmidt rank ($R$) after a single-particle measurement can only take the values $R \in \{1, 2\}$, and the topological vocabulary of \textit{linked} versus \textit{unlinked} rings is correspondingly binary.\\

\noindent The present work extends this correspondence to the system of qutrits. Moving from $d = 2$ to $d = 3$ (where $d$ representing the dimension of the system) immediately breaks the binary vocabulary that made the qubit correspondence so clean, as we discuss in Sec.~\ref{sec:qutrit-obstruction}, the residual pair left behind after measuring one particle of a tripartite qutrit system can now be \textit{fully separated} ($R = 1$), \textit{partially entangled} ($R = 2$), or \textit{maximally entangled} ($R = 3$). A single ring, which can only be linked or unlinked, cannot encode this three-valued structure. We resolve this obstruction by replacing each topological ring with a cable of $n = d - 1 = 2$ parallel strands (Sec.~\ref{sec:qutrit-obstruction}), so that the number of strands ($n$) remaining linked after a measurement, $n = 0$ (no strand connected), $n = 1$ (one strand connected), or $n = 2$ (two strands connected), reproduces the three possible values of $R$ exactly. This cabling device is deliberately treated as a labeling convention rather than a proven topological invariant. The precise sense in which it should and should not be interpreted is discussed explicitly in Sec.~\ref{sec:status-of-correspondence}.\\

\noindent Equipped with this extended dictionary, we analyze two families of tripartite qutrit states under both computational-basis(CB) and mutually unbiased basis (MUB) measurement of a single particle. Sec.~\ref{sec:ghz-states} treats the qutrit $GHZ$ state $|GHZ_3\rangle$, and shows that it reproduces the clean qubit-level picture in full: every CB outcome yields $R = 1$, and every MUB outcome yields the maximal rank $R = 3$, with no intermediate case ever realized (full derivations shown in Appendix~\ref{app:ghz-measurement}). Sec.~\ref{sec:w-states} turns to the qutrit $W$ class states. Because the Hilbert space dimension grows from $2^3 = 8$ to $3^3 = 27$ in passing from qubits to qutrits, the single qubit $W$ state fragments into a family of symmetric states built from \textit{permutation bags} of indices drawn from $\{0, 1,2\}$;  six $\ket{W_{p,p,q}^{sym}}$ states and one $\ket{W_{\{0,1,2\}}}$ state. We show that these two categories behave quite differently under measurement (full derivations in Appendices~\ref{app:wstates}--\ref{app:w-mub-alldifferent}): the \textit{two same-one different} family yields a probability-weighted mixture of $R = 1$ and $R = 2$ under CB measurement, while the \textit{all-different} state yields $R = 2$ uniformly. Under MUB measurement, the \textit{two same - one different} family is confined to $R = 2$ with an unequal Schmidt spectrum, while the \textit{all-different} state reaches the maximal rank $R = 3$. Sec.~\ref{sec:comparative-structure} brings these results together and applies the \textit{cabled-link dictionary} to them directly, introducing a \emph{hybrid} classification for the cases, realized only by the \textit{two same - one different} $W$ states under CB measurement --- in which the residual link type is not uniform across measurement outcomes: a predominant link, realized with the higher outcome probability, together with a contextual exception realized at the remaining outcome. Table~\ref{tab:topological-summary} summarizes the resulting classification for every state and basis treated in this work. The central structural finding of this work is that the relevant divide among the states studied is not $GHZ$ class versus $W$ class, but \textit{repeated-index} versus \textit{all-different-index} bag structure: $\ket{GHZ_3}$ and the \textit{all-different} $W$ state $\ket{W_{\{0,1,2\}}}$ share a clean, outcome-independent splitting pattern in both bases, while the six $\ket{W_{p,p,q}^{sym}}$ states are the only states in this work that require a hybrid, probability-weighted classification.

\section{Preliminaries}
\label{sec:preliminaries}

\subsection{Qutrit Computational Basis and Mutually Unbiased Bases}

\noindent A single qutrit is a three-level quantum system defined on the Hilbert space $\mathcal{H}=\mathbb{C}^3$. Its standard CB is the orthonormal set given as
\begin{equation}
\label{cb1}
B_0 = \{\ket{0}, \ket{1}, \ket{2}\}, \qquad \langle i | k \rangle = \delta_{ik}.
\end{equation}
A tripartite qutrit system resides in the tensor product space $\mathcal{H}_A \otimes \mathcal{H}_B \otimes \mathcal{H}_C \cong \mathbb{C}^{27}$, with the CB naturally extending to $\ket{a}_A \ket{b}_B \ket{c}_C = \ket{abc}$, where $a,b,c \in \{0,1,2\}$.

\subsubsection{Mutually Unbiased Bases (MUBs)} 
\noindent Two orthonormal bases $\{\ket{a_i}\}$ and $\{\ket{b_j}\}$ of a $d$-dimensional Hilbert space are \emph{mutually unbiased} if the magnitude of the inner product between any two basis vectors is strictly uniform:
\begin{equation}
\label{eq:mub-def}
|\langle a_i | b_j \rangle|^2 = \frac{1}{d}, \qquad \text{for every } i,j.
\end{equation}
Physically, this means a state prepared in one basis yields a completely random, uniform outcome distribution when measured in the other. For a prime dimension $d$, the maximum number of pairwise Mutually Unbiased Bases (MUB) is exactly $d+1$\cite{wkmb1989,tbikmb2010}. For a qutrit system where $d=3$, which is a prime number, this maximum number of MUB is $4$, where one is our known CB (see Eq.~(\ref{cb1})). However, the rest of them will be non-computational bases (NCB). 

\paragraph{Explicit Construction :} 

\noindent We construct this complete set explicitly, as measuring particle $A$ in these NCB is central to uncovering the hidden topological entanglement signatures. These are shown in the appendices.

\noindent Let $\omega=e^{2\pi \ii/3}$ be the primitive cube root of unity, satisfying $\omega^3=1$ and $1+\omega+\omega^2=0$. For $m,j \in \{0,1,2\}$, we define the NCB vectors as
\begin{equation}
\label{eq:ncb}
\ket{e_j^{(m)}} = \frac{1}{\sqrt{3}} \sum_{k=0}^{2} \omega^{jk+mk^2} \ket{k}
\end{equation}
Together with the CB $B_0$, the three constructed NCB bases $B_{m+1} = \{\ket{e_0^{(m)}}, \ket{e_1^{(m)}}, \ket{e_2^{(m)}}\}$ for $m \in \{0,1,2\}$ constitute the complete set of $4$ MUBs for a qutrit. 

\paragraph{Properties of the Cube Root of Unity :}
To prove that these constructed bases of Eq.~(\ref{eq:ncb}) are both orthonormal and mutually unbiased, we rely on a fundamental property of the root of unity. For any integer $n$, the sum of the powers of $\omega$ forms a geometric series:
\begin{equation}
\label{eq:omega-sum}
\sum_{k=0}^{2} \omega^{nk} = 
\begin{cases} 
3 & \text{if } n \equiv 0 \pmod 3 \\ 
0 & \text{if } n \not\equiv 0 \pmod 3 
\end{cases}
\end{equation}
If $n$ is a multiple of $3$, then $\omega^n = 1$, and every term in the sum is $1$, yielding $3$. If $n$ is not a multiple of $3$, then $\omega^n \neq 1$, and evaluating the finite geometric series we get 
\begin{equation}
\label{omega3}
\frac{(\omega^n)^3 - 1}{\omega^n - 1} = \frac{1-1}{\omega^n - 1} = 0.
\end{equation}

\paragraph{Proof of Orthonormality\\}

To verify orthonormality, we compute the inner product between two arbitrary vectors
$|e_j^{(m)}\rangle$ and $|e_{j'}^{(m)}\rangle$ belonging to the same basis. Taking the Hermitian conjugate of the ket gives

\begin{equation}
\langle e_j^{(m)}|
=
\frac{1}{\sqrt3}
\sum_{k=0}^{2}
\omega^{-(jk+mk^2)}
\langle k|.
\end{equation}

\noindent Therefore,

\begin{align}
\langle e_j^{(m)}
|
e_{j'}^{(m)}
\rangle
&=
\left(
\frac1{\sqrt3}
\sum_{k=0}^{2}
\omega^{-(jk+mk^2)}
\langle k|
\right)
\left(
\frac1{\sqrt3}
\sum_{\ell=0}^{2}
\omega^{j'\ell+m\ell^2}
|\ell\rangle
\right)
\nonumber\\
&=
\frac13
\sum_{k=0}^{2}
\sum_{\ell=0}^{2}
\omega^{-(jk+mk^2)}
\omega^{j'\ell+m\ell^2}
\langle k|\ell\rangle.
\end{align}

\noindent Since the computational basis is orthonormal,

\begin{equation}
\langle k|\ell\rangle
=
\delta_{k\ell},
\end{equation}

\noindent only the terms with $k=\ell$ survive, giving

\begin{align}
\langle e_j^{(m)}
|
e_{j'}^{(m)}
\rangle
&=
\frac13
\sum_{k=0}^{2}
\omega^{-(jk+mk^2)}
\omega^{j'k+mk^2}.
\end{align}

\noindent Next, we combine the exponents:

\begin{align}
-(jk+mk^2)+(j'k+mk^2)
&=
-jk-mk^2+j'k+mk^2 \nonumber
\\
&=
(j'-j)k.
\end{align}

\noindent It is to be noted that the quadratic terms $mk^2$ cancel exactly because both vectors belong to the same basis and therefore have the same value of $m$.

\noindent Hence,

\begin{equation}
\label{eq:orthogonality}
\boxed{
\langle e_j^{(m)}
|
e_{j'}^{(m)}
\rangle
=
\frac13
\sum_{k=0}^{2}
\omega^{(j'-j)k}
}.
\end{equation}

\noindent We now use the identity given in Eq.~(\ref{eq:omega-sum}). There are two possible cases.

\paragraph{Case 1: $j=j'$.}

\noindent In this case,
\begin{equation}
j'-j=0,
\end{equation}

\noindent so that

\begin{equation}
\sum_{k=0}^{2}
\omega^{0k}
=
\sum_{k=0}^{2}1
=
3.
\end{equation}

\noindent Substituting into Eq.~\eqref{eq:orthogonality},

\begin{equation}
\langle e_j^{(m)}
|
e_j^{(m)}
\rangle
=
\frac13(3)
=
1.
\end{equation}

\noindent Thus every basis vector is normalized.

\paragraph{Case 2: $j\neq j'$.}

Since $j,j'\in\{0,1,2\}$,

\begin{equation}
j'-j
\equiv
1
\quad\text{or}\quad
2
\pmod3.
\end{equation}

\noindent Therefore,

\begin{equation}
\sum_{k=0}^{2}
\omega^{(j'-j)k}
=
0,
\end{equation}

\noindent and hence

\begin{equation}
\langle e_j^{(m)}
|
e_{j'}^{(m)}
\rangle
=
0.
\end{equation}

\noindent Thus distinct vectors are orthogonal.\\

\noindent Combining both cases, we conclude that
\begin{equation}
\boxed{
\langle e_j^{(m)}
|
e_{j'}^{(m)}
\rangle
=
\delta_{jj'}.
}
\end{equation}

\noindent This in turn proves that every constructed basis
\begin{equation}
B_{m+1}
=
\{
|e_0^{(m)}\rangle,
|e_1^{(m)}\rangle,
|e_2^{(m)}\rangle
\}
\end{equation}
is orthonormal.

\noindent 

\paragraph{Proof of Mutual Unbiasedness :}

To prove that the four bases $B_0,\;B_1,\;B_2,\;B_3$ form a complete set of mutually unbiased bases, we must show that the squared magnitude of the inner product between any vector belonging to one basis and any vector belonging to a different basis is exactly $\frac{1}{3}$.\\

\textit{Unbiasedness with the Computational Basis}

The given computational basis is $B_0$ (Eq.~(\ref{cb1}), while the remaining three bases are generated from Eq.(\ref{eq:ncb}).
We first calculate the overlap between a computational basis vector $\ket{a}$ and one of the constructed vectors. Using the orthonormality of the computational basis,

\begin{align}
\langle a|e_j^{(m)}\rangle
&=
\left\langle a\left|
\frac1{\sqrt3}
\sum_{k=0}^{2}
\omega^{jk+mk^2}\ket{k}
\right.\right\rangle
\nonumber\\[1ex]
&=
\frac1{\sqrt3}
\sum_{k=0}^{2}
\omega^{jk+mk^2}
\langle a|k\rangle
\nonumber\\[1ex]
&=
\frac1{\sqrt3}
\omega^{ja+ma^2},~~~~~\text{as}~~\langle a|k\rangle=\delta_{ak}.
\end{align}

\noindent Since every power of $\omega$ has unit modulus, $|\omega^{ja+ma^2}|=1$.\\

\noindent Therefore,

\begin{equation}
\boxed{
|\langle a|e_j^{(m)}\rangle|^2
=
\frac13.
}
\end{equation}

\noindent Hence every constructed basis is mutually unbiased with respect to the computational basis.\\

\textit{Unbiasedness Between Two Constructed Bases}

\noindent Now consider two different constructed bases, say, $B_{m+1} \qquad\text{and}\qquad B_{m'+1}$, where $m\neq m'$. Take two arbitrary basis vectors $\ket{e_j^{(m)}}, \quad \ket{e_{j'}^{(m')}}$. The first vector has the expansion

\begin{equation}
\ket{e_j^{(m)}}
=
\frac1{\sqrt3}
\sum_{k=0}^{2}
\omega^{jk+mk^2}\ket{k},
\end{equation}

\noindent whose Hermitian conjugate is

\begin{equation}
\bra{e_j^{(m)}}
=
\frac1{\sqrt3}
\sum_{k=0}^{2}
\omega^{-(jk+mk^2)}
\bra{k},
\end{equation}

\noindent since complex conjugation changes $\omega^n \longrightarrow \omega^{-n}$. Similarly, we have

\begin{equation}
\ket{e_{j'}^{(m')}}
=
\frac1{\sqrt3}
\sum_{\ell=0}^{2}
\omega^{j'\ell+m'\ell^2}
\ket{\ell}.
\end{equation}

\noindent Their inner product is therefore

\begin{align}
\langle e_j^{(m)}
|
e_{j'}^{(m')}
\rangle
&=
\left(
\frac1{\sqrt3}
\sum_{k=0}^{2}
\omega^{-(jk+mk^2)}
\bra{k}
\right)
\left(
\frac1{\sqrt3}
\sum_{\ell=0}^{2}
\omega^{j'\ell+m'\ell^2}
\ket{\ell}
\right)
\nonumber\\[1ex]
&=
\frac13
\sum_{k=0}^{2}
\sum_{\ell=0}^{2}
\omega^{-(jk+mk^2)}
\omega^{j'\ell+m'\ell^2}
\langle k|\ell\rangle.
\end{align}

\noindent Since the computational basis is orthonormal, $\langle k|\ell\rangle = \delta_{k\ell}$, all terms vanish except those satisfying $k=\ell$. Hence the double summation reduces to

\begin{align}
\langle e_j^{(m)}
|
e_{j'}^{(m')}
\rangle
&=
\frac13
\sum_{k=0}^{2}
\omega^{-(jk+mk^2)}
\omega^{j'k+m'k^2}.
\end{align}

\noindent Combining the exponents gives

\begin{align}
-(jk+mk^2)+(j'k+m'k^2)
&=
-jk-mk^2+j'k+m'k^2
\nonumber\\
&=
(j'-j)k+(m'-m)k^2.
\end{align}

\noindent Introducing the shorthand notation $\Delta j=j'-j, \qquad \Delta m=m'-m$, the inner product becomes

\begin{equation}
\langle e_j^{(m)}
|
e_{j'}^{(m')}
\rangle
=
\frac13
\sum_{k=0}^{2}
\omega^{\Delta jk+\Delta mk^2}.
\end{equation}

\noindent Since the two bases are different, $m\neq m'$, we necessarily have $\Delta m\in\{1,2\} \pmod3$.\\

\textit{Evaluating the Sum\\} 
The summation contains only the three values $k=0,1,2$. For $k=0$, $\Delta jk+\Delta mk^2=0$, so the first term is $1$. For $k=1$, the second term is $\Delta j+\Delta m$. For $k=2$, the third term is $2\Delta j+4\Delta m$ (since $4\equiv1\pmod 3$, this becomes $2\Delta j+\Delta m$). Therefore, the sum becomes

\begin{equation}
S
=
1
+
\omega^{\Delta j+\Delta m}
+
\omega^{2\Delta j+\Delta m},
\end{equation}

\noindent and consequently we get

\begin{equation}
\langle e_j^{(m)}
|
e_{j'}^{(m')}
\rangle
=
\frac13S.
\end{equation}

\noindent We now evaluate the six possible values of $(\Delta j,\Delta m)$. Since arithmetic is performed modulo $3$, we have
\begin{equation}
\Delta j\in\{0,1,2\},
\qquad
\Delta m\in\{1,2\},
\end{equation}
\noindent because $m\neq m'$.


\paragraph{Case I : $\Delta m=1$}

\noindent There are three possible values of $\Delta j$.

\begin{itemize}

\item $\Delta j=0$:

\begin{equation}
S
=
1+\omega^{0+1}+\omega^{0+1}
=
1+2\omega.
\end{equation}

\item $\Delta j=1$:

\begin{align}
S
&=
1+\omega^{1+1}+\omega^{2+1}\nonumber\\
&=
1+\omega^2+\omega^3\nonumber\\
&=
1+\omega^2+1\nonumber\\
&=
2+\omega^2.
\end{align}

\item $\Delta j=2$:

\begin{align}
S
&=
1+\omega^{2+1}+\omega^{4+1}\nonumber\\
&=
1+\omega^3+\omega^5\nonumber\\
&=
1+1+\omega^2\nonumber\\
&=
2+\omega^2,
\end{align}

\noindent since
\begin{equation}
\omega^5=\omega^{3+2}=\omega^2.
\end{equation}

\end{itemize}


\paragraph{Case II : $\Delta m=2$}

\noindent Again there are three possibilities.

\begin{itemize}

\item $\Delta j=0$:

\begin{equation}
S
=
1+\omega^{0+2}+\omega^{0+2}
=
1+2\omega^2.
\end{equation}

\item $\Delta j=1$:

\begin{align}
S
&=
1+\omega^{1+2}+\omega^{2+2}\nonumber\\
&=
1+\omega^3+\omega^4\nonumber\\
&=
1+1+\omega\nonumber\\
&=
2+\omega,
\end{align}

\noindent because
\begin{equation}
\omega^4=\omega^{3+1}=\omega.
\end{equation}

\item $\Delta j=2$:

\begin{align}
S
&=
1+\omega^{2+2}+\omega^{4+2}\nonumber\\
&=
1+\omega^4+\omega^6\nonumber\\
&=
1+\omega+1\nonumber\\
&=
2+\omega,
\end{align}

\noindent since
\begin{equation}
\omega^4=\omega,
\qquad
\omega^6=(\omega^3)^2=1.
\end{equation}

\end{itemize}


\noindent Thus every possible value of $S$ reduces to one of the four complex numbers, $1+2\omega,\quad 1+2\omega^2,\quad 2+\omega,\quad 2+\omega^2$. To compute their magnitudes, we use the fact that

\begin{equation}
\omega^2=\overline{\omega},
\qquad
\omega+\overline{\omega}
=
2\cos\left(\frac{2\pi}{3}\right)
=
-1.
\end{equation}

\noindent For any integers $a$ and $b$,

\begin{align}
|a+b\omega|^2
&=
(a+b\omega)(a+b\overline{\omega})\nonumber\\
&=
a^2+ab(\omega+\overline{\omega})+b^2\nonumber\\
&=
a^2-ab+b^2.
\end{align}

\noindent Similarly,

\begin{equation}
|a+b\omega^2|^2
=
a^2-ab+b^2.
\end{equation}

\noindent Applying this formula, $|1+2\omega|^2 = 1-2+4 =3$, $|1+2\omega^2|^2= 3$, $|2+\omega|^2= 4-2+1 =3$, $|2+\omega^2|^2 = 3$. Hence, in every case, $|S|^2=3$. Substituting this into the inner product,

\begin{align}
\left|
\langle e_j^{(m)}
|
e_{j'}^{(m')}
\rangle
\right|^2
&=
\frac19|S|^2\nonumber\\
&=
\frac19(3)\nonumber\\
&=
\frac13.
\end{align}

\noindent Therefore,

\begin{equation}
\boxed{
\left|
\langle e_j^{(m)}
|
e_{j'}^{(m')}
\rangle
\right|^2
=
\frac13,
\qquad
m\neq m',
}
\end{equation}

\noindent showing that every pair of distinct constructed bases is mutually unbiased.

\paragraph{The Inverse Relation :}

\noindent Since each constructed basis $B_{m+1} = \left\{ \ket{e_0^{(m)}},\ket{e_1^{(m)}},\ket{e_2^{(m)}} \right\}$ has been shown to be orthonormal, it forms a complete basis for the three-dimensional Hilbert space. Consequently, it satisfies the
completeness relation

\begin{equation}
\label{eq:completeness}
\sum_{j=0}^{2}
\ket{e_j^{(m)}}
\bra{e_j^{(m)}}
=
I,
\end{equation}

\noindent where $I$ denotes the identity operator on the qutrit Hilbert space.\\

\noindent To express the computational basis (CB) vectors in terms of the MUB vectors, we act with the identity operator on an arbitrary computational basis state $\ket{a}$, where $a\in\{0,1,2\}$ such that

\begin{align}
\ket{a}
&=
I\ket{a}
\nonumber\\
&=
\left(
\sum_{j=0}^{2}
\ket{e_j^{(m)}}
\bra{e_j^{(m)}}
\right)
\ket{a}
\nonumber\\
&=
\sum_{j=0}^{2}
\ket{e_j^{(m)}}
\langle e_j^{(m)}|a\rangle.
\label{eq:inverse1}
\end{align}

\noindent Thus, the only quantity that remains to be determined is the overlap $\langle e_j^{(m)}|a\rangle$.\\


\noindent From the definition of the constructed basis vectors of Eq.(\ref{eq:ncb}), we first compute

\begin{align}
\langle a|e_j^{(m)}\rangle
&=
\left\langle
a
\left|
\frac1{\sqrt3}
\sum_{k=0}^{2}
\omega^{jk+mk^2}
\ket{k}
\right.
\right\rangle
\nonumber\\
&=
\frac1{\sqrt3}
\sum_{k=0}^{2}
\omega^{jk+mk^2}
\langle a|k\rangle
\nonumber\\
&=
\frac1{\sqrt3}
\omega^{ja+ma^2},
\end{align}

\noindent since

\begin{equation}
\langle a|k\rangle
=
\delta_{ak}.
\end{equation}

\noindent Taking the complex conjugate gives

\begin{align}
\langle e_j^{(m)}|a\rangle
&=
\overline{\langle a|e_j^{(m)}\rangle}
\nonumber\\
&=
\frac1{\sqrt3}
\omega^{-(ja+ma^2)},
\end{align}

\noindent because

\begin{equation}
\overline{\omega^n}
=
\omega^{-n}.
\end{equation}

\noindent Substituting this result into Eq.~(\ref{eq:inverse1}) gives

\begin{align}
\ket{a}
&=
\sum_{j=0}^{2}
\ket{e_j^{(m)}}
\left(
\frac1{\sqrt3}
\omega^{-(ja+ma^2)}
\right)
\nonumber\\
&=
\frac1{\sqrt3}
\sum_{j=0}^{2}
\omega^{-(ja+ma^2)}
\ket{e_j^{(m)}},
\qquad
a\in\{0,1,2\}.
\label{eq:inverse-mub-prelim}
\end{align}

\noindent Thus every computational basis vector can be written as a linear combination of the vectors belonging to any one of the mutually unbiased bases.

\noindent Since every coefficient is an integer power of the primitive cube root of unity $\omega$, each coefficient has unit modulus. Consequently, the inverse transformation differs from the forward transformation only by the complex-conjugation of the phase factors. This inverse expansion is used repeatedly throughout this work to rewrite tripartite qutrit states whenever particle $A$ is measured in a non-computational basis.

\subsection{Schmidt Decomposition and Schmidt Rank}
\label{sec:schmidt-decomp}

\noindent The Schmidt decomposition is a standard tool of quantum information theory \cite{terhal2000,kumar2024}. Any normalized pure state of a bipartite system $B\otimes C$ (where each is a qutrit) can be written generally as

\begin{equation}
\label{schmidt}
\ket{\psi}_{BC} = \sum_{j,k=0}^{2} M_{jk}\ket{j}_B\ket{k}_C.
\end{equation}

\noindent The coefficients $M_{jk}$ completely define a $3\times3$ complex matrix $M$, with rows indexed by particle $B$ and columns by particle $C$. By applying the singular value decomposition \cite{klema1980,baker2005}, $M=U\Sigma V^\dagger$, we can rotate the local bases of $B$ and $C$ to rewrite the state in the Schmidt decomposition
\begin{equation}
\label{schmidtrank}
\ket{\psi}_{BC} = \sum_{i=0}^{2}\sigma_i\ket{\tilde{\imath}}_B\ket{\tilde{\imath}}_C,
\end{equation}
where $\sigma_i \ge 0$ and $\sum_i \sigma_i^2 = 1$.\\

\noindent The \emph{Schmidt rank} $R$ is simply the number of strictly non-zero $\sigma_i$ terms. Physically, $R=1$ indicates that $\ket{\psi}_{BC}$ is a fully separable product state, while $R>1$ signals the survival of bipartite entanglement. The residual qutrit pair reaches \emph{maximal entanglement} if all the non-zero Schmidt coefficients are perfectly equal.\\

\noindent The squared Schmidt coefficients $\lambda_i=\sigma_i^2$ are exactly the eigenvalues of the (unnormalized) reduced density matrix $\rho_B \propto M^\dagger M$ (or equivalently $M M^\dagger$). Therefore, we can find the Schmidt rank purely by analyzing the coefficient matrix $M$ without needing to perform a full singular value decomposition.\\

\noindent In the appendices, for every post-measurement state, we compute:
\begin{equation}
\label{cardinal}
R = \#\{\text{non-zero eigenvalues of } M^\dagger M\}, \qquad \{\lambda_i\} = \mathrm{eig}(M^\dagger M),
\end{equation}
\noindent where $\#$ is the cardinality of the set.\\

\noindent \emph{Note on diagonal matrices:} If the coefficient matrix $M$ naturally takes a diagonal form, then $M^\dagger M$ is also diagonal. In these instances, the eigenvalues $\{\lambda_i\}$ are simply the squared magnitudes of the diagonal entries, and the characteristic equation requires no further algebraic solving.

\subsection{Topological Interpretation of Entanglement}
\subsubsection{Topological Links}
\label{sec:topological-links}

\noindent We use standard definitions from knot theory \cite{rolf1976} to classify the entanglement structure of the states studied in this work:

\begin{description}
\item \emph{Unlink:} A collection of loops that are not topologically connected; each loop can be separated from the others without any cutting.
\item \emph{Borromean Rings:} A set of three rings in which no two are pairwise linked, yet the three together cannot be separated: like a Brunnian configuration\footnote{A Brunnian configuration (or Brunnian link) is a set of closed loops that are linked together. They cannot come apart as a whole group. Yet, if you cut or remove any single loop, the remaining loops fall completely apart.}. Cutting any one ring releases the other two.
\item \emph{Hopf Link:} The simplest non-trivial link between two rings: a pair of rings that are linked and cannot be separated without cutting one of them.
\item \emph{$n$-Hopf Link:} A generalization of the Hopf link to $n$ components, in which every component is pairwise linked to every other component. Cutting any one component leaves the remaining $(n-1)$ components still robustly linked to each other.
\end{description}

\noindent These four link types anchor the correspondence developed in the remainder of this section: the Borromean rings and the Hopf link give the qubit-level dictionary of Aravind's original proposal (Sec.~\ref{aravinds}), and the $n$-Hopf link with $n=3$ gives the \textbf{full cabling} endpoint of the qutrit generalization introduced in Sec.~\ref{sec:qutrit-obstruction}.

\subsubsection{Aravind's Correspondence for Qubits} 
\label{aravinds}
\begin{figure}[H]
    \centering
    \begin{subfigure}[b]{0.45\textwidth}
        \centering
        \includegraphics[width=\textwidth]{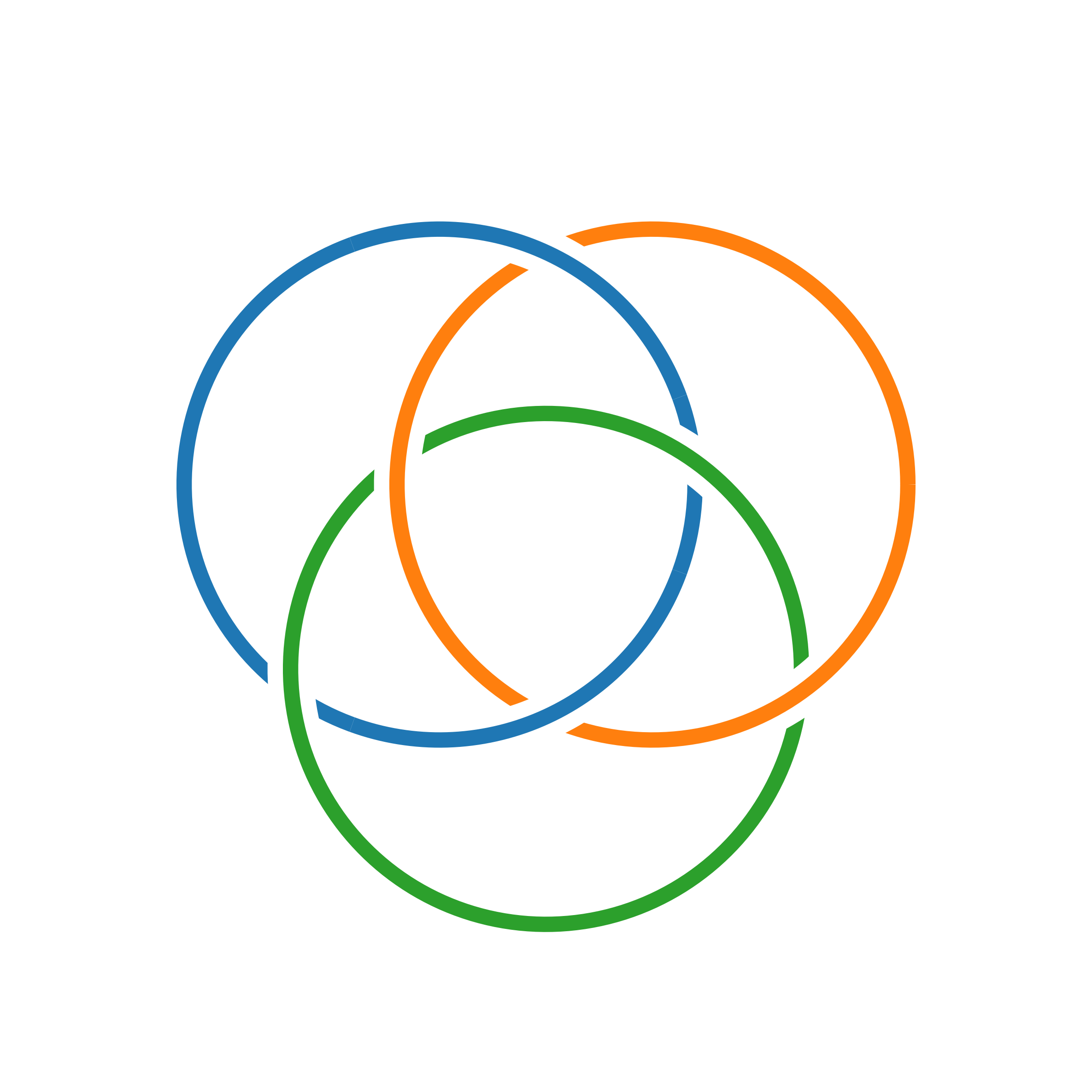}
        \caption{Borromean rings}
        \label{fig:qubit_borromean}
    \end{subfigure}
    \hfill
    \begin{subfigure}[b]{0.45\textwidth}
        \centering
        \includegraphics[width=\textwidth]{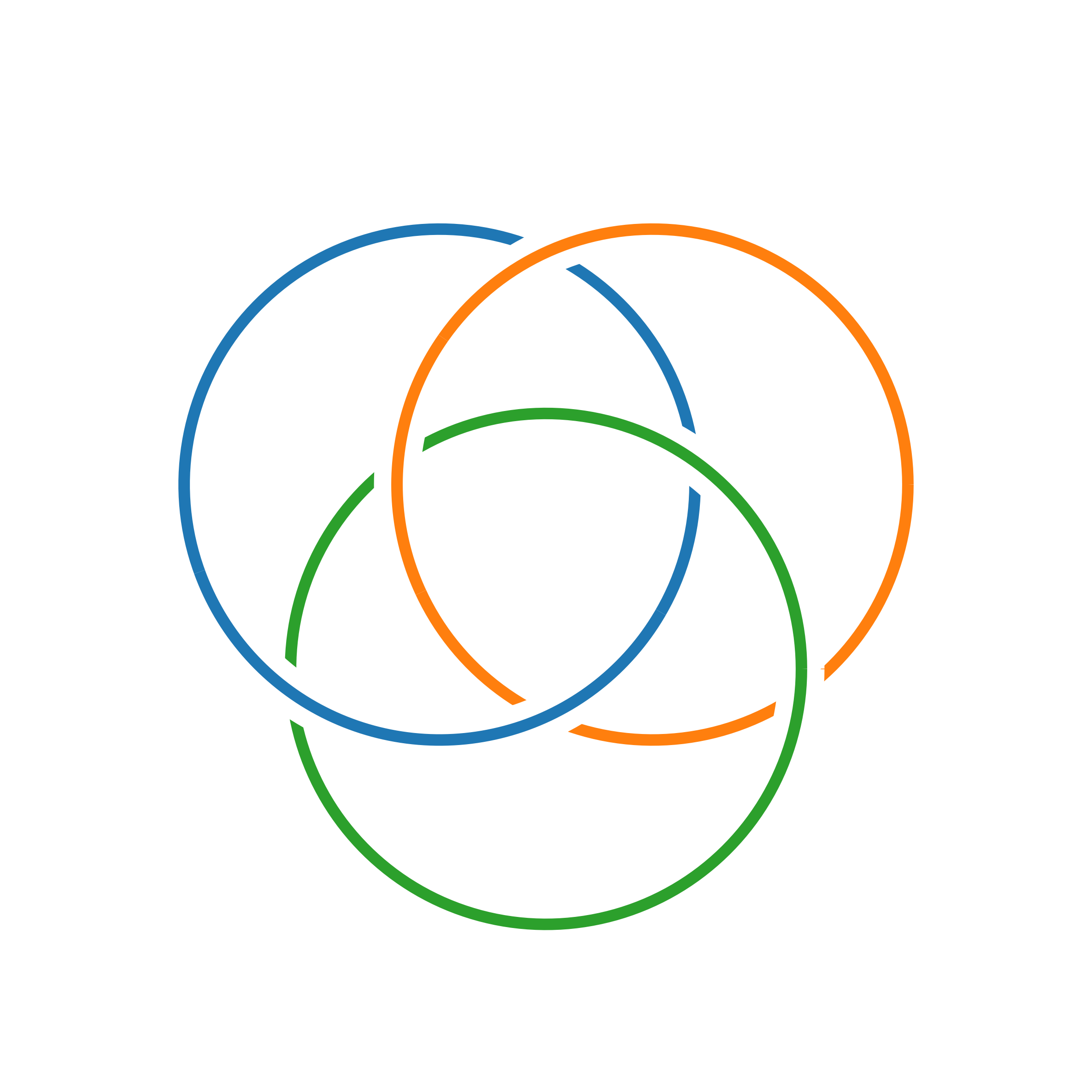}
        \caption{Three-Hopf Ring}
        \label{fig:qubit_hopf}
    \end{subfigure}
    \caption{The two qubit-level link types behind Aravind's correspondence. (a) The Borromean Rings: no two rings are linked on their own, but cutting any one still frees the other two completely. (b) The 3-Hopf Ring: each ring is linked to its neighbor, so cutting one ring still leaves the other two linked to each other.}
    \label{fig:qubit_baseline}
\end{figure}

\noindent For multi-qubit systems, Aravind elegantly proposed a direct topological dictionary mapping quantum states to physical knots and links. In this framework, a single qubit is represented by a closed physical ring, and a multi-qubit entangled state is represented by an interlinked configuration of these rings \cite{aravind1997}. Crucially, the act of performing a projective measurement on a qubit \cite{sougata2509} (thereby removing it from the combined system) corresponds to physically \emph{cutting} its respective ring and pulling it out of the configuration (See Figs.~(\ref{fig:qubit_borromean}) and (\ref{fig:qubit_hopf})). \\

\subsubsection{The Qutrit Obstruction and Two-Strand Cabling}
\label{sec:qutrit-obstruction} 
\noindent This binary vocabulary (linked vs. unlinked) works perfectly for qubits, where the residual Schmidt rank can only be $R \in \{1,2\}$. However, when measuring one particle of a tripartite qutrit system, the remaining pair has three possible states viz. separable ($R=1$), partially entangled ($R=2$), or maximally entangled ($R=3$). \\

\noindent To capture this higher-dimensional structure faithfully, we extend Aravind's framework by replacing single topological rings with \emph{cables} consisting of two parallel strands ($n=d-1$), a construction familiar from the classical theory of cable knots \cite{adams1994}.\\

\begin{figure}[H]
    \centering
    \begin{subfigure}[b]{0.45\textwidth}
        \centering
        \includegraphics[width=\textwidth]{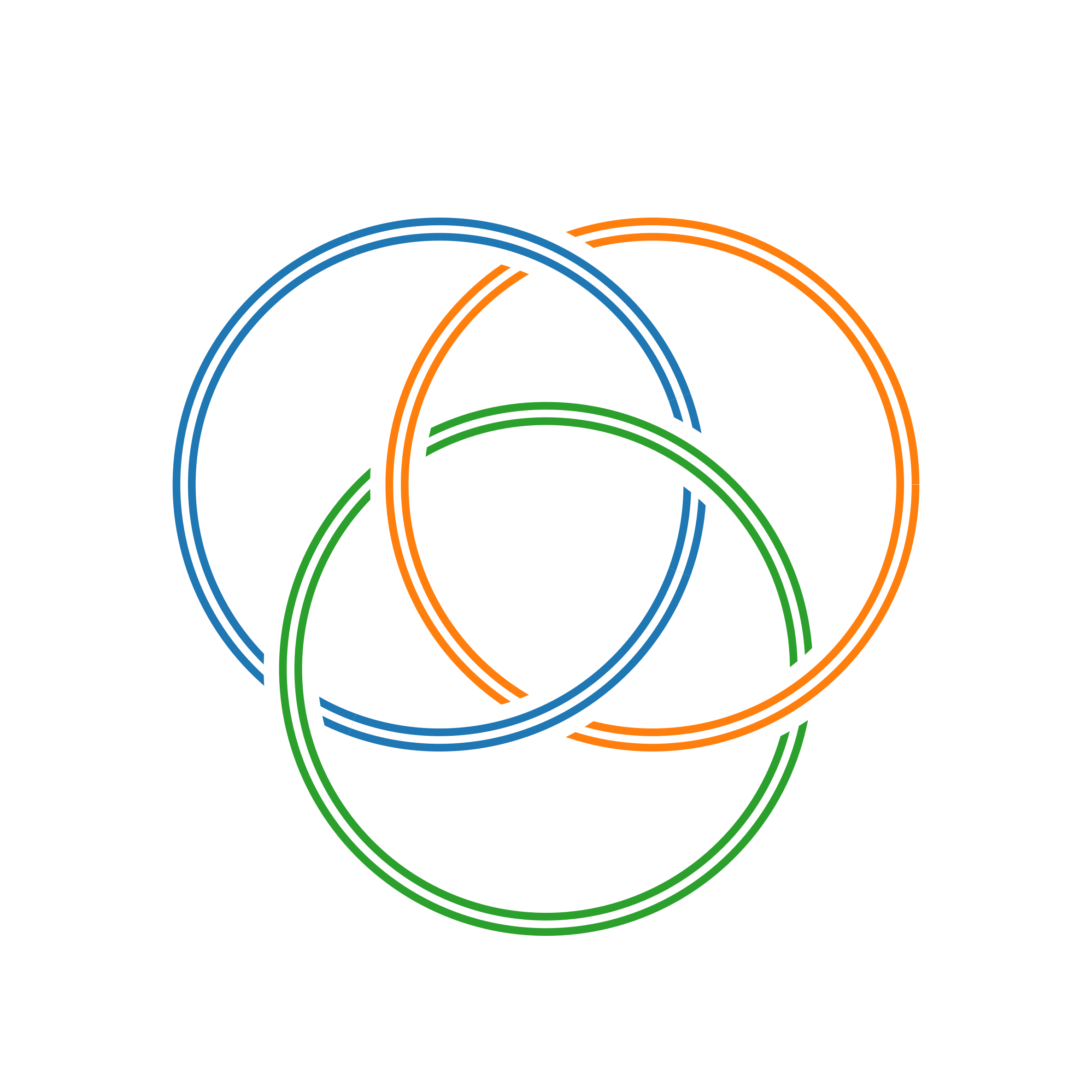}
        \caption{Cabled Borromean link}
        \label{fig:qutrit_borromean}
    \end{subfigure}
    \hfill
    \begin{subfigure}[b]{0.45\textwidth}
        \centering
        \includegraphics[width=\textwidth]{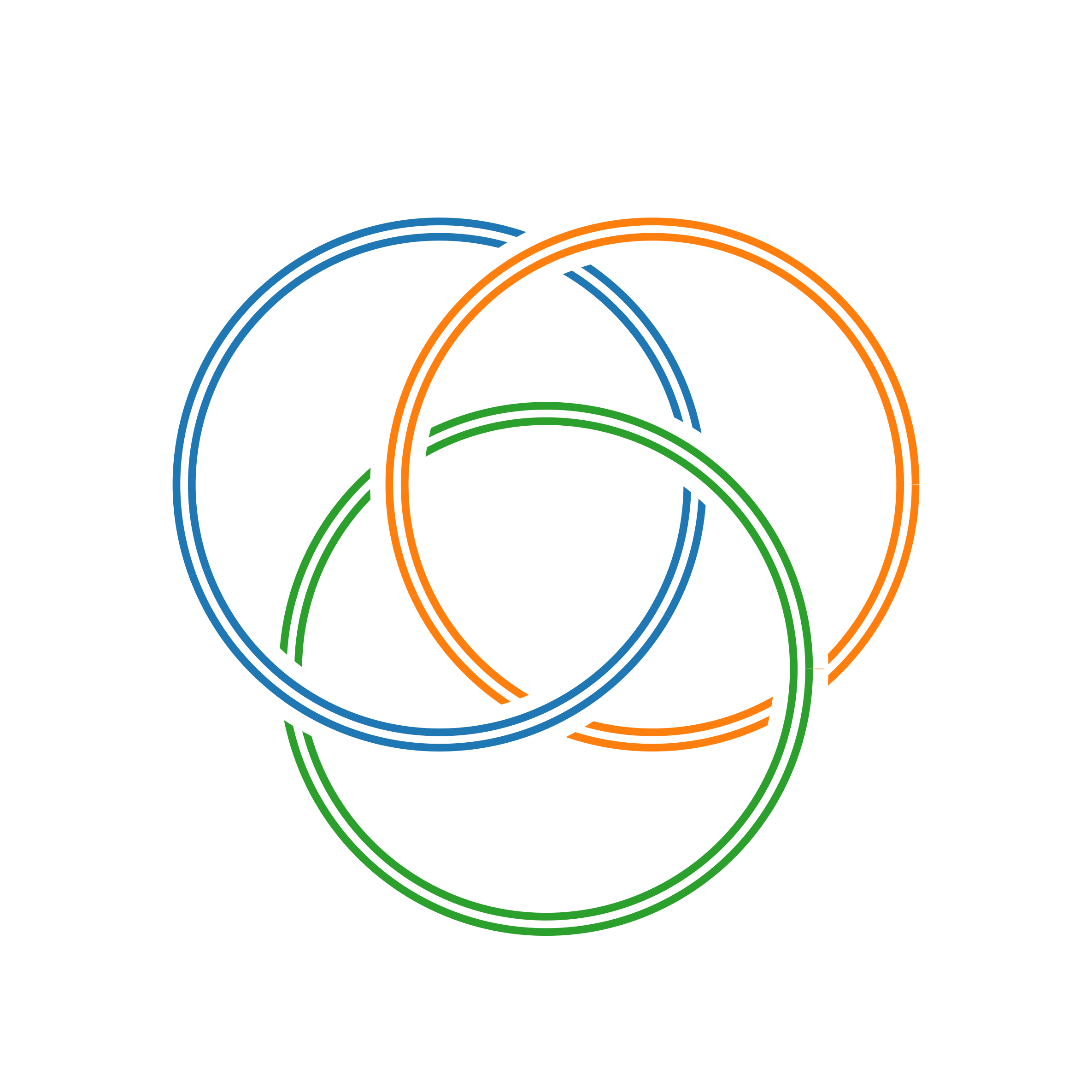}
        \caption{Cabled 3-Hopf link}
        \label{fig:qutrit_hopf}
    \end{subfigure}
    \caption{Each ring from Figure~\ref{fig:qubit_baseline} is now drawn as a cable of $n=d-1=2$ parallel strands, one extra strand for the one extra level a qutrit has over a qubit. How many of these strands stay linked after a measurement cut is what tells the link types apart at the qutrit level.}
    \label{fig:qutrit_cabling}
\end{figure}

\begin{figure}[H]
    \centering
    \begin{subfigure}[b]{0.45\textwidth}
        \centering
        \includegraphics[width=\textwidth]{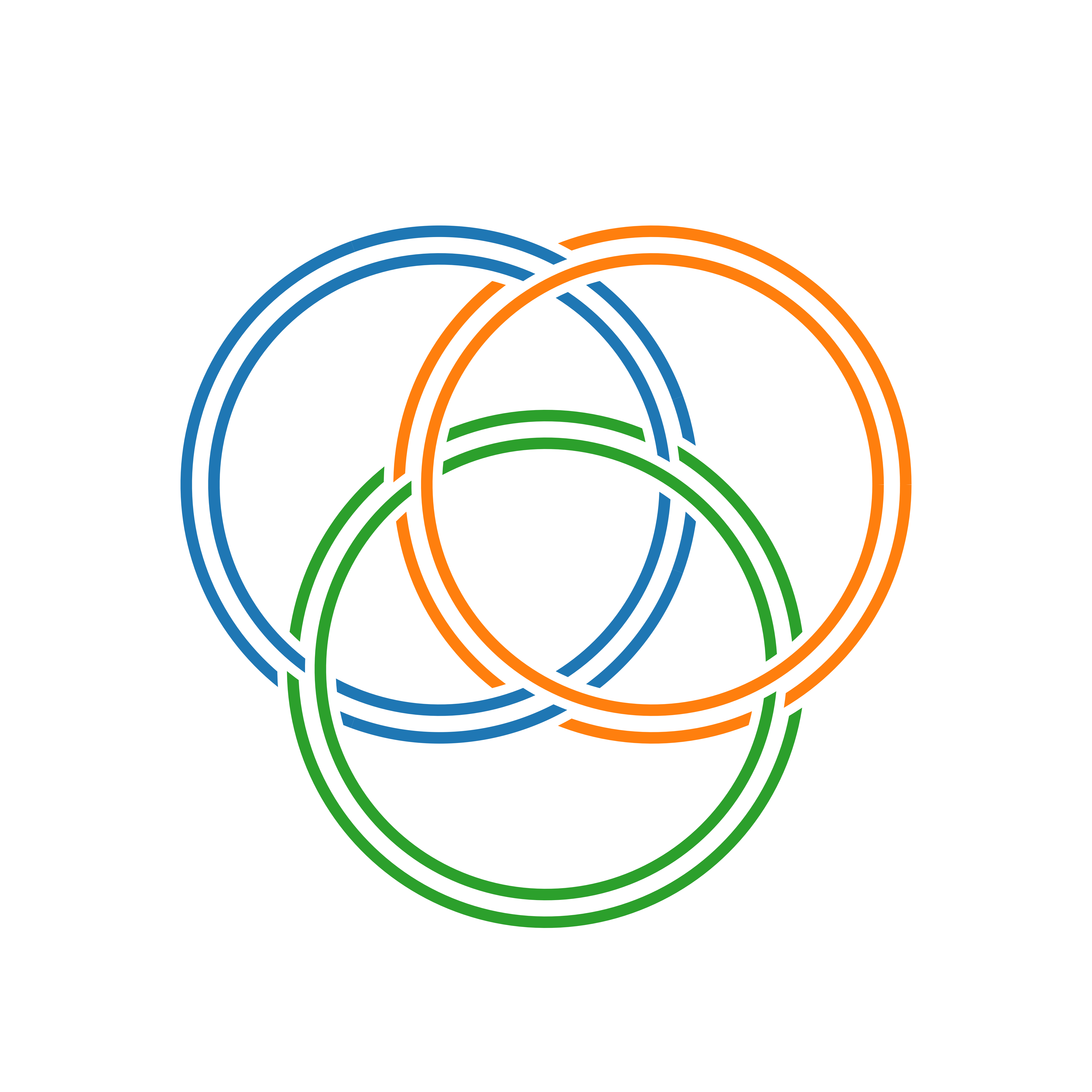}
        \caption{Partially cabled, Borromean side}
        \label{fig:partial_borromean}
    \end{subfigure}
    \hfill
    \begin{subfigure}[b]{0.45\textwidth}
        \centering
        \includegraphics[width=\textwidth]{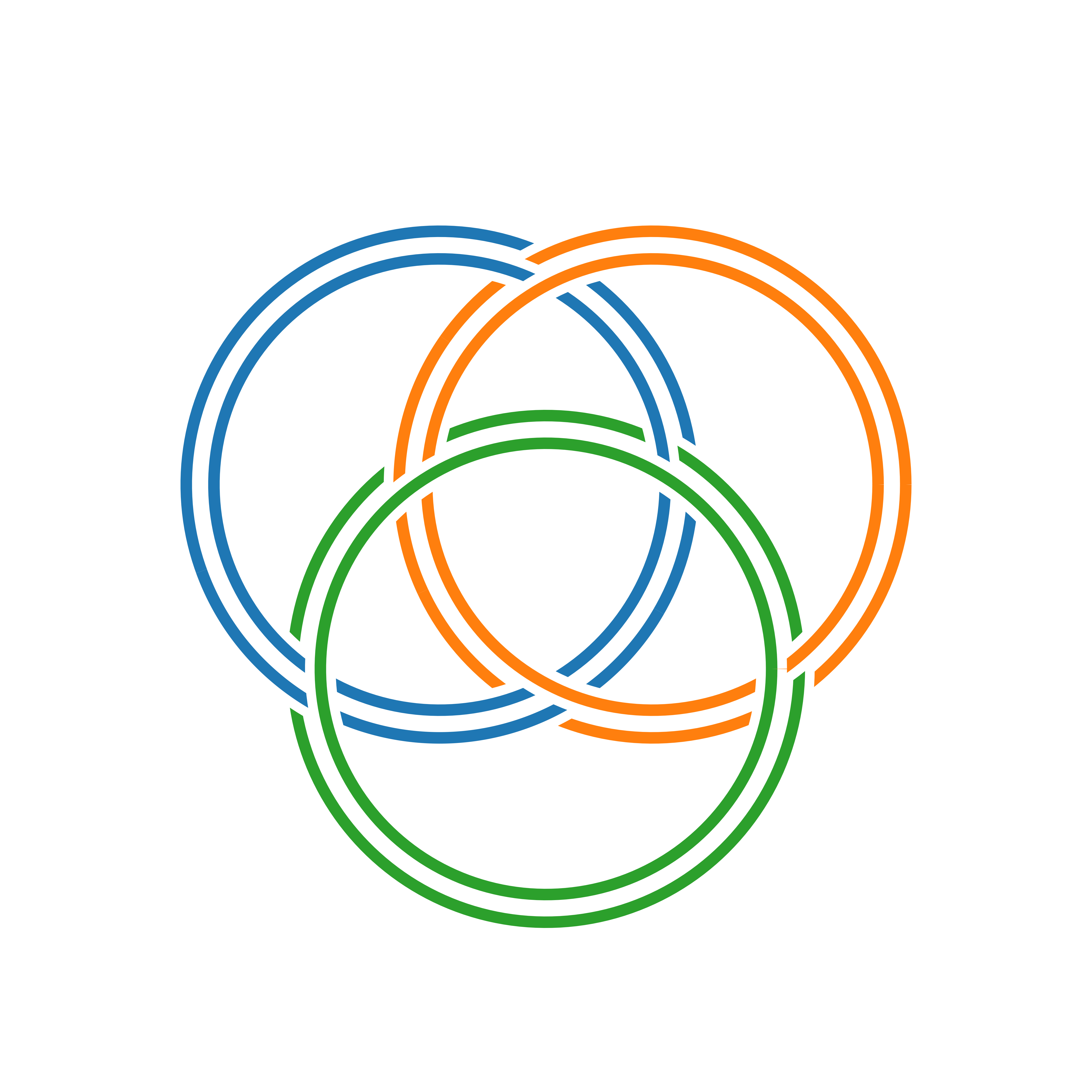}
        \caption{Partially cabled, Hopf side}
        \label{fig:partial_hopf}
    \end{subfigure}
    \caption{The $R=2$ case: one strand of the cable stays linked between the two remaining rings, while the other is cut and plays no further part. Only one of the two strands is doing any topological work here, exactly what \textit{partial cabling} means.}
    \label{fig:partial_cabling}
\end{figure}

\begin{figure}[H]
    \centering
    \begin{subfigure}[b]{\textwidth}
        \centering
        \includegraphics[width=\textwidth]{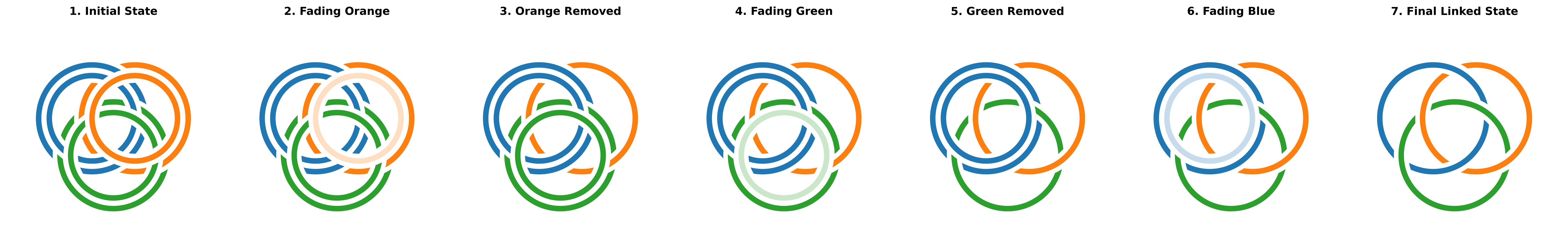}
        \caption{Borromean side}
        \label{fig:seq_borromean}
    \end{subfigure}

    \vspace{1.5em}

    \begin{subfigure}[b]{\textwidth}
        \centering
        \includegraphics[width=\textwidth]{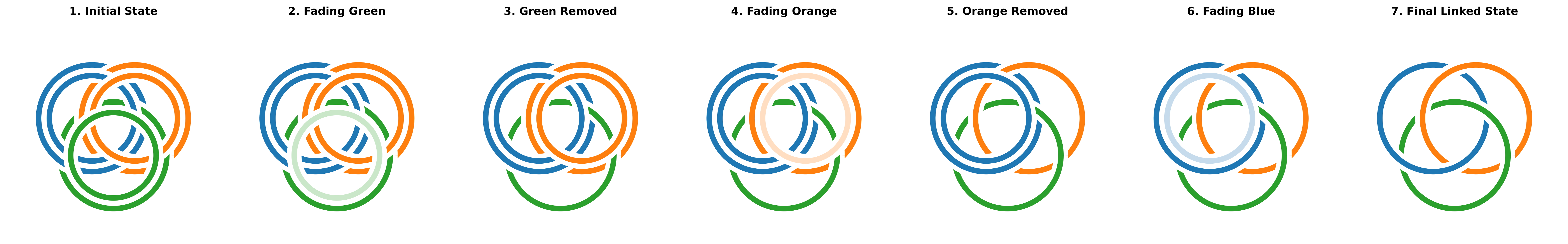}
        \caption{Hopf side}
        \label{fig:seq_hopf}
    \end{subfigure}
    \caption{How Fig.~\ref{fig:partial_cabling} actually comes about: starting from the fully cabled tangle, one strand after another loosens and fades out, until only a single strand is left holding each pair of rings together. The rings are still genuinely linked one connecting strand is enough for that but now only one strand is doing the work, with its former partner hanging loose. This is the $R=2$, partial-cabling case, shown as it is built rather than as a finished picture.}
    \label{fig:seq_partial}
\end{figure}

\noindent When evaluating the residual entanglement between two cabled rings after a measurement cut, there are exactly three topological outcomes for the remaining strands, which map one-to-one with the Schmidt rank:
\begin{itemize}
    \item \textit{$R=1 \leftrightarrow$ Fully Separated (0 strands linked):} The generalized Borromean (GHZ) behavior. The measurement completely severs the correlation.
    \item \textit{$R=2 \leftrightarrow$ Partial Cabling (1 strand linked):} Entanglement survives, but only partially -- not at its strongest.
    \item \textit{$R=3 \leftrightarrow$ Full Cabling (2 strands linked):} The generalized 3-Hopf behavior. The maximum possible bipartite entanglement is robustly preserved.
\end{itemize}

\noindent This extended vocabulary directly matches the integer count of the Schmidt rank to the integer count of topologically linked strands.

\begin{figure}[H]
    \centering
    \includegraphics[width=\textwidth]{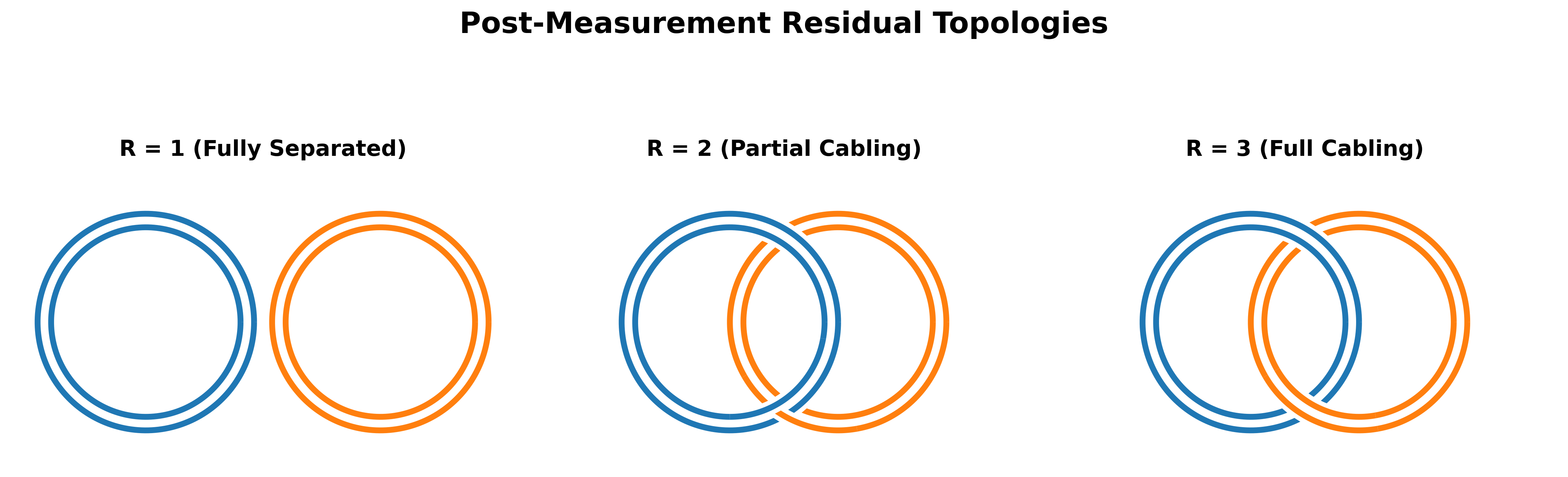}
    \caption{The full dictionary in one picture: the number of strands still linked between the two remaining rings exactly matches the Schmidt rank $R$ left behind after measurement, fully separated for $R=1$, one strand linked for $R=2$, both strands linked for $R=3$.}
    \label{fig:residual_strands}
\end{figure}

\subsubsection{Status of the Correspondence}
\label{sec:status-of-correspondence}

\noindent The \textit{cabling dictionary} introduced above is a descriptive bookkeeping device, not a derived topological invariant. It is important to be precise about what has and has not been established.\\

\noindent The construction proceeds in one direction only. We first compute the Schmidt rank $R$ of the residual state algebraically, via the eigenvalues of $M^\dagger M$, and only afterward assign it a \textit{cabled-link} picture with $R-1$ linked strands. No independent topological quantity, a linking number, a braid-group representation, a polynomial invariant such as the Jones or Kauffman bracket, is computed from the link diagrams and shown to reproduce $R$. The correspondence is therefore built to match the data by construction. It has not been shown to constrain or predict anything beyond what the coefficient-matrix calculation already gives.\\

\noindent Relatedly, \textit{the cutting} operation used throughout, removing a ring or a strand to model a projective measurement, has no counterpart in the quantum mechanical formalism itself. Projective measurement is a linear map on the Hilbert space while strand removal is an operation on a classical link diagram. Aravind's original correspondence for qubits was explicit that this identification is an analogy motivated by a shared \emph{qualitative} feature (both operations can destroy or preserve a \textit{notion of connectedness}), not an isomorphism between two mathematical structures, a distinction examined in detail elsewhere \cite{kauffman2011}, and the same caveat applies here with added force: the two-strand cabling was chosen specifically because $n = d-1$ reproduces the right count of intermediate ranks for a qutrit, not because any physical or topological argument singles out that choice of $n$. The choice of a two-strand cable is therefore not uniquely singled out by any known topological invariant or physical principle. Rather, it is adopted as a convenient representational device whose sole purpose is to distinguish the three possible residual Schmidt-rank values, $R \in \{1,2,3\}$. Consequently, any alternative labeling framework that preserves this one-to-one distinction would be equally adequate for the descriptive purposes of the present correspondence.\\

\noindent This basis-dependence is not a new concern specific to our qutrit extension. Even at the qubit level, Sugita showed that Aravind's original \textit{GHZ-Borromean identification} depends on the choice of measurement basis, and proposed reformulating the correspondence in a basis-independent way through the reduced density matrix rather than through a fixed link diagram \cite{sugita2007}. The \textit{cabling dictionary} developed here inherits the same basis-dependence at the qutrit level: the link assignments in Secs~\ref{sec:ghz-states} and~\ref{sec:w-states} are computed for the computational basis and the mutually unbiased bases specifically, and nothing in the construction guarantees that an analogous dictionary would hold, or would need reformulating, for a generic unlisted basis.\\

\noindent Two further limitations are worth stating plainly. First, the dictionary has only been checked against the \emph{permutation-symmetric} $GHZ$ and $W$ class states treated in this paper; nothing here establishes that it extends to the full space of SLOCC-inequivalent three-qutrit entanglement classes, which is considerably larger than the two families considered. Second, the dictionary is defined only for pure states and residual bipartite Schmidt rank; no attempt is made here to extend it to mixed states or to a genuinely multi-partite (rather than bipartite-after-measurement) topological quantity.\\

\noindent None of this undermines the usefulness of the dictionary as an organizing language for the results of Secs.~\ref{sec:ghz-states}---\ref{sec:comparative-structure}: it gives a compact and consistent vocabulary for describing how entanglement redistributes under measurement, and the hybrid classification of Table~\ref{tab:topological-summary} is meaningful independent of any topological reading. But it should be read as a naming convention layered on top of the Schmidt-rank calculation, not as evidence that the entanglement structure of these states is \textit{topological} in any stronger sense.

\section{$GHZ$-Class States in Qutrit Systems}
\label{sec:ghz-states}

\subsection{Definition of the Qutrit GHZ States}

\noindent The qutrit generalization of the three-qubit GHZ state \cite{GHZ1989,ghz32018} is defined as the maximally-correlated superposition over all three computational basis values appearing identically on all three particles:
\begin{equation}
\label{ghzstate}
\ket{GHZ_3} = \frac{1}{\sqrt3}\big(\ket{000}+\ket{111}+\ket{222}\big) = \frac{1}{\sqrt3}\sum_{a=0}^{2}\ket{a}_A\ket{a}_B\ket{a}_C.
\end{equation}
Like its qubit counterpart, $\ket{GHZ_3}$ is fully permutation-symmetric under the exchange of particles $A$, $B$, and $C$. Because of this symmetry, measuring any single particle yields mathematically equivalent results for the residual pair. Therefore, without loss of generality, we designate particle $A$ for all projective measurements.

\subsection{Measurement-Induced Entanglement Splitting}

\noindent When measuring particle $A$ in the computational basis $B_0$, each possible outcome $j\in\{0,1,2\}$ occurs with a uniform probability of $p(j)=\frac{1}{3}$. This measurement collapses the residual pair onto the separable product state $\ket{j}_B\ket{j}_C$, resulting in a Schmidt rank of $R=1$ with no surviving entanglement. Conversely, if particle $A$ is measured in any of the three non-computational MUBs ($B_1, B_2, B_3$), every outcome again occurs with probability $p(j)=\frac{1}{3}$. However, the residual pair is now left in the maximally entangled state $\frac{1}{\sqrt3}\big[\varepsilon_{j,0}\ket{00}+\varepsilon_{j,1}\ket{11}+\varepsilon_{j,2}\ket{22}\big]$. This resulting state possesses a maximal Schmidt rank of $R=3$, characterized by perfectly equal eigenvalues $\{\lambda_i\}=\{\frac{1}{3},\frac{1}{3},\frac{1}{3}\}$. Full derivations for both cases are given in Appendix~\ref{app:ghz-measurement}.\\

\noindent Table~\ref{tab:main-ghz-summary} summarizes these splitting behaviors. The measurement outcomes are uniform across all choices of $j$ and simultaneously across all three MUBs. Consequently, $\ket{GHZ_3}$ exhibits a sharp topological dichotomy: the CB acts as a \textit{cutting} basis that completely destroys entanglement, while the MUBs act as \textit{linking} bases that preserve entanglement at its absolute maximum, with no intermediate ranks realized.

\begin{table}[H]
\centering
\caption{Measurement-induced entanglement splitting for $\ket{GHZ_3}$; particle $A$ measured in the computational basis $B_0$ or in any non-computational MUB $B_1,B_2,B_3$ ($m=0,1,2$).}
\renewcommand{\arraystretch}{1.4}
\begin{tabular}{c|c|c|c}
\toprule
Basis measured on $A$ & $p(j)$, $j=0,1,2$ & Eigenvalues $\{\lambda_i\}$ & $R$ \\
\midrule
$B_0$ (computational) & $\frac{1}{3}$ & $\{1,0,0\}$ & 1 \\
Any MUB ($B_1,B_2,B_3$) & $\frac{1}{3}$ & $\left\{\tfrac13,\tfrac13,\tfrac13\right\}$ & 3 \\
\bottomrule
\end{tabular}
\label{tab:main-ghz-summary}
\end{table}

\section{$W$-Class States in Qutrit Systems}
\label{sec:w-states}
 
\subsection{Definition of the Symmetric Qutrit $W$-States}
 
\noindent To extend the topological analysis of entanglement splitting from qubits to qutrits, we must first establish the appropriate $d=3$ analogues of the $W$ state. In standard three-qubit systems, the canonical $W$ state \cite{Dur2000} is an equal superposition of all permutations of one excitation shared among three particles, defined by perfect permutation symmetry and the survival of bipartite entanglement upon the loss of a single particle. However, moving from a qubit system (states $\ket{0}, \ket{1}$) to a qutrit system (states $\ket{0}, \ket{1}, \ket{2}$) significantly expands the Hilbert space dimension from $2^3 = 8$ to $3^3 = 27$. Consequently, the concept of a \textit{single $W$ state} fragments into a broader family of symmetric states. To systematically construct this family, we utilize the concept of \textit{permutation bags—multisets} of three indices drawn from $\{0, 1, 2\}$. A generalized qutrit $W$ state is formed by taking an equal superposition of all unique permutations of the indices within a given \textit{bag}. Because we require non-trivial entanglement, excluding fully separable states like $\ket{000}$, these \textit{bags} naturally fall into two distinct structural categories.\\
 
\paragraph{Category 1: The \textit{Two Same, One Different} States :}

\noindent The most direct structural translation of the standard qubit $W$ state involves \textit{bags} where two particles share one state, and the third particle holds a different state. There are exactly \textit{six possible bags} of this type in a qutrit system, leading to six distinct symmetric states. Each state comprises three permutations.
\begin{itemize}
    \item The first two states represent the distribution of a single higher-energy excitation ($\ket{1}$ or $\ket{2}$) against a ground-state background ($\ket{0}$):
        \begin{itemize}
            \item Bag $\{0,0,1\}$: $\ket{W_{1}^{sym}} = \frac{1}{\sqrt{3}}(\ket{001}+\ket{010}+\ket{100})$.
            \item Bag $\{0,0,2\}$: $\ket{W_{2}^{sym}} = \frac{1}{\sqrt{3}}(\ket{002}+\ket{020}+\ket{200})$.
        \end{itemize}
    \item Two further states mirror the inverted $W$ state, where one particle is in the ground state while two are excited:
        \begin{itemize}
            \item Bag $\{1,1,0\}$: $\ket{W_{3}^{sym}} = \frac{1}{\sqrt{3}}(\ket{110}+\ket{101}+\ket{011})$.
            \item Bag $\{2,2,0\}$: $\ket{W_{5}^{sym}} = \frac{1}{\sqrt{3}}(\ket{220}+\ket{202}+\ket{022})$.
        \end{itemize}
    \item The remaining two states describe interactions entirely between the excited levels, ignoring the ground state:
        \begin{itemize}
            \item Bag $\{1,1,2\}$: $\ket{W_{4}^{sym}} = \frac{1}{\sqrt{3}}(\ket{112}+\ket{121}+\ket{211})$.
            \item Bag $\{2,2,1\}$: $\ket{W_{6}^{sym}} = \frac{1}{\sqrt{3}}(\ket{221}+\ket{212}+\ket{122})$.
        \end{itemize}
\end{itemize}
 
\paragraph{Category 2: The \textit{All Different} State :}

\noindent To fully utilize the three-dimensional nature of the qutrit space while maintaining $W$-type symmetry, we must consider the case where all three particles occupy distinct states. There is only one such bag: $\{0,1,2\}$. Because all three indices are unique, this bag generates $3! = 6$ distinct permutations. The resulting state is the most dimensionally complete symmetric $W$-state in the tripartite qutrit regime:
\begin{equation}
\label{allw}
\ket{W_{\{0,1,2\}}} = \frac{1}{\sqrt{6}}( \ket{012} + \ket{021} + \ket{102} + \ket{120} + \ket{201} + \ket{210} ).
\end{equation}
 
\noindent By establishing this comprehensive family of states, we can rigorously analyze how symmetric entanglement splits under measurement across different bases.
 
\subsection{Measurement-Induced Entanglement Splitting}
 
\paragraph{Computational Basis :}

\noindent When measuring particle $A$ in the CB $B_0$, the \textit{two-same one-different} states (i.e. $\ket{W_{p,p,q}^{sym}}$) do not yield uniform outcomes. For a state derived from bag $\{p,p,q\}$, outcome $p$ occurs with probability $\frac{2}{3}$ and leaves the residual $B$-$C$ pair in the maximally entangled qubit-pair state $\frac{1}{\sqrt2}(\ket{pq}+\ket{qp})$, with $R=2$. Outcome $q$ occurs with probability $\frac{1}{3}$ and yields the separable product state $\ket{pp}$ with $R=1$, while the third computational outcome never occurs. In contrast, the \textit{all-different} state (i.e. $\ket{W_{\{0,1,2\}}}$) yields a uniform probability of $\frac{1}{3}$ for every outcome. In all cases for the \textit{all-different} (i.e. $\ket{W_{\{0,1,2\}}}$) state, the residual pair retains a Schmidt rank of $R=2$ with eigenvalues $\{\frac{1}{2}, \frac{1}{2}, 0\}$; entangled, but not maximally so despite utilizing all three levels. Full derivations are given in Appendices~\ref{app:wstates}, \ref{app:w012}, and the complete table in Appendix~\ref{app:comp-summary}.
 
\paragraph{Mutually Unbiased Basis :}

\noindent Measuring particle $A$ in any of the non-computational MUBs ($B_1, B_2, B_3$) yields a perfectly uniform probability $p(j)=\frac{1}{3}$ for every outcome across both state categories. The \textit{two same - one different} states ($\ket{W_{p,p,q}^{sym}}$) retain a Schmidt rank of $R=2$, but their residual eigenvalues become unequal: $\{\frac{3+\sqrt5}{6}, \frac{3-\sqrt5}{6}, 0\}$. The \textit{all-different} state ($\ket{W_{\{0,1,2\}}}$) elevates to a full Schmidt rank of $R=3$, but its spectrum is similarly unequal with eigenvalues $\{\frac{2}{3}, \frac{1}{6}, \frac{1}{6}\}$. Full derivations are given in Appendices~\ref{app:mub-wstates}, \ref{app:mub-w012}, and the complete table in Appendix~\ref{app:mub-summary}.\\
 
\noindent Table~\ref{tab:main-w-summary} collects the splitting behaviors across both bases. Unlike the $GHZ$ state, the $W$-class states do not exhibit a clean topological dichotomy: the $\ket{W_{p,p,q}^{sym}}$ family yields a probability-weighted mixture of ranks under standard measurement, and every MUB measurement produces an unequal Schmidt spectrum rather than a perfectly flat one.
 
\begin{table}[H]
\centering
\caption{Measurement-induced entanglement splitting for the symmetric qutrit $W$-states; particle $A$ measured in the computational basis $B_0$ or in any non-computational MUB ($m=0,1,2$).}
\renewcommand{\arraystretch}{1.5}
\begin{tabular}{c|c|c|c|c}
\toprule
State family & Basis & $p(j)$ & Eigenvalues $\{\lambda_i\}$ & $R$ \\
\midrule
\multirow{3}{*}{\shortstack{$\ket{W_{p,p,q}^{sym}}$}} & $B_0$: outcome $p$ & $\frac{2}{3}$ & $\{\tfrac12,\tfrac12,0\}$ & $2$ \\
 & $B_0$: outcome $q$ & $\frac{1}{3}$ & $\{1,0,0\}$ & $1$ \\
 & Any MUB & $\frac{1}{3}$ & $\left\{\tfrac{3+\sqrt5}{6},\tfrac{3-\sqrt5}{6},0\right\}$ & $2$ \\
\midrule
\multirow{2}{*}{$\ket{W_{\{0,1,2\}}}$} & $B_0$ & $\frac{1}{3}$ & $\{\tfrac12,\tfrac12,0\}$ & $2$ \\
 & Any MUB & $\frac{1}{3}$ & $\left\{\tfrac23,\tfrac16,\tfrac16\right\}$ & $3$ \\
\bottomrule
\end{tabular}
\label{tab:main-w-summary}
\end{table}

\section{Comparative Topological Structure of $GHZ$- and $W$-Class States}
\label{sec:comparative-structure}
 
\noindent The Preliminaries established a dictionary between Schmidt rank and \textit{cabled-link} type such as (a) \textit{Borromean-type} ($R=1$), (b) \textit{partial cabling} ($R=2$) and (c) \textit{full cabling} / $3$-Hopf-type ($R=3$), extending an approach we have also used to classify measurement-induced entanglement splitting across other tripartite state families \cite{sougata2509}. A single named link can be assigned outright to an entire measurement basis when the residual Schmidt rank is the \emph{same for every outcome} $j$. When it is not, we report the \emph{predominant link}, the one realized with the higher outcome probability, together with the \emph{contextual exception} realized at the remaining outcome(s), and describe the basis as a hybrid of the two.\\

\noindent We now apply this dictionary to the results of Secs.~\ref{sec:ghz-states} and ~\ref{sec:w-states}, checking outcome-uniformity explicitly in each case, and use the comparison to draw out the structural difference between the $GHZ$- and $W$-class states.
 
\subsection{$\ket{GHZ_3}$: The Dictionary in Its Cleanest Form}
 
\noindent Table~\ref{tab:main-ghz-summary} shows that both of $\ket{GHZ_3}$'s measurement bases pass the uniformity test outright: every outcome $j=0,1,2$ gives $R=1$ under $B_0$, and every outcome gives $R=3$ under any of $B_1,B_2,B_3$. A single link can therefore be assigned to each basis without qualification.

\begin{itemize}
\item $B_0\to$ \textit{Borromean-type (generalized)}: both strands cut.
\item Any MUB $\to$ \textit{full cabling / 3-Hopf-type (generalized)}: no strands cut.
\end{itemize}

\noindent $\ket{GHZ_3}$ is thus the case, the cabling dictionary was built for: (i) two bases, two links, no intermediate case realized and (ii) no ambiguity in either assignment. Every other state in this paper deviates from this picture in at least one of the two bases.
 
\subsection{The $\ket{W_{p,p,q}^{sym}}$-States: A Hybrid Topological Entity}
 
\paragraph{Computational Basis (Predominant Partial Cabling, Contextual Borromean Exception):}

\noindent Table~\ref{tab:main-w-summary} shows that $\ket{W_{p,p,q}^{sym}}$ under $B_0$ is \emph{not} uniform: outcome $p$ (probability $\tfrac23$) gives $R=2$, matching \textit{partial cabling} (one strand cut); outcome $q$ (probability $\tfrac13$) gives $R=1$, matching \textit{Borromean-type} (both strands cut); the third computational value never occurs at all.\\

\noindent Following the hybrid-link convention above, we describe $B_0$ for this family as predominantly \textit{partial cabling}, with a contextual Borromean exception at the minority outcome $q$.\\

\paragraph{Mutually Unbiased Basis (A Clean Link) :} 

\noindent Under any MUB, every outcome of $\ket{W_{p,p,q}^{sym}}$ gives $p(j)=\tfrac13$ and $R=2$ uniformly, so \textit{partial cabling} can be assigned to $B_1,B_2,B_3$ without ambiguity: one strand cut, one intact, for every outcome and every choice of MUB.
 
\subsection{The $\ket{W_{\{0,1,2\}}}$ State: Clean Ranks Throughout}
 
\noindent Table~\ref{tab:main-w-summary} shows that $\ket{W_{\{0,1,2\}}}$ passes the uniformity test in \emph{both} bases, every outcome occurs with probability $\tfrac13$ under $B_0$ and under any MUB, so a single clean link can be assigned in both cases, unlike the $\ket{W_{p,p,q}^{sym}}$ family above.
\begin{itemize}
\item $B_0\to$ \textit{partial cabling} ($R=2$): one strand cut, one intact -- and uniform, unlike the $\ket{W_{p,p,q}^{sym}}$'s $B_0$ result above.
\item Any MUB $\to$ \textit{full cabling / 3-Hopf-type} ($R=3$): no strands cut, matching $\ket{GHZ_3}$'s MUB result.
\end{itemize}

\subsection{Summary}
\begin{table}[H]
\centering
\caption{Topological classification of every state and basis treated in this work.}
\renewcommand{\arraystretch}{1.5}
\footnotesize
\begin{tabular}{c|c|c|c}
\toprule
State & Basis & Uniform across $j$? & Link \\
\midrule
\multirow{2}{*}{$\ket{GHZ_3}$} & $B_0$ & Yes & Borromean\\
 & Any MUB & Yes & 3-Hopf \\
\midrule
\multirow{3}{*}{$\ket{W_{p,p,q}^{sym}}$} & $B_0$: outcome $p$ (predom., $\frac{2}{3}$) & No & Partial cabling \\
 & $B_0$: outcome $q$ (context., $\frac{1}{3}$) & No & Borromean\\
 & Any MUB & Yes & Partial cabling \\
\midrule
\multirow{2}{*}{$\ket{W_{\{0,1,2\}}}$} & $B_0$ & Yes & Partial cabling \\
 & Any MUB & Yes & 3-Hopf \\
\bottomrule
\end{tabular}
\label{tab:topological-summary}
\end{table}
 
\noindent The comparison in Table~\ref{tab:topological-summary} is the central structural finding of this work. $\ket{GHZ_3}$ realizes the cabling dictionary in its purest form: two bases, two link types, uniform outcomes throughout. The $\ket{W_{\{0,1,2\}}}$ state matches this same clean pattern in both of its bases. Only the $\ket{W_{p,p,q}^{sym}}$ family's computational basis departs from outcome-uniformity, and even there, the departure is not a failure of classification but a hybrid topological entity; a predominant link realized with probability $\tfrac23$, and a contextual exception realized with probability $\tfrac13$.

\section{Results and Discussion}
\label{sec:results-discussion}

\noindent The measurement-induced entanglement splitting derived in Sections~\ref{sec:ghz-states} and~\ref{sec:w-states}, and classified topologically in Section~\ref{sec:comparative-structure}, supports three observations that go beyond the individual calculations.

\paragraph{The GHZ state is the special case, not the template.}
\noindent Table~\ref{tab:topological-summary} shows that $|GHZ_3\rangle$ is the only state in this work for which \emph{both} measurement bases pass the outcome-uniformity test outright. This is a direct consequence of the fact that $|GHZ_3\rangle$ is built from a single computational \textit{bag} $\{a,a,a\}$ repeated identically across all three particles. Every outcome $j$ on
particle $A$ leaves an identical residual structure on $B$ and $C$, up to relabeling. The $W$-class states, by contrast, are built from bags with nontrivial index structures --- either a repeated-index pattern $\{p,p,q\}$ or an all-different pattern $\{0,1,2\}$ --- and it is exactly this asymmetry in the bag structure that is responsible for the non-uniform, hybrid behavior found for the two-same-one-different family under $B_0$ (Section~\ref{sec:w-states}, Appendix~\ref{app:comp-summary}). The clean topological dictionary of Section~\ref{sec:qutrit-obstruction} should therefore be understood as applying naturally to permutation-symmetric states with non-repeating bag structure, of which $|GHZ_3\rangle$ and $|W_{\{0,1,2\}}\rangle$ are the two examples treated here, rather than to permutation-symmetric states in general.

\paragraph{The all-different $W$-state is topologically closer to GHZ than to the other
$W$-states.}
\noindent This is, in our view, the most interesting structural result of the paper. Despite belonging to the $W$-class by construction --- a single shared excitation pattern, permutation symmetry, survival of bipartite entanglement upon particle loss --- $|W_{\{0,1,2\}}\rangle$ behaves under the cabling dictionary exactly as $|GHZ_3\rangle$ does: a single, unambiguous link per
basis, uniform across all three outcomes (Sec.~\ref{sec:comparative-structure}). The six  $\ket{W_{p,p,q}^{sym}}$ states are the only ones that require the hybrid classification at all. This suggests that, at least as far as the topological dictionary is sensitive to it, the relevant structural divide among the qutrit $W$-class states is not $W$-class versus $GHZ$-class but rather \emph{repeated-index} versus \emph{all-different-index} bags, with the latter inheriting the clean behavior usually associated with $GHZ$-type states. We regard this as an empirical pattern established by direct calculation (Tables~\ref{tab:main-ghz-summary} and~\ref{tab:main-w-summary}) rather than as something explained by the \textit{cabling picture} itself (see the caveats below).

\paragraph{What the hybrid classification does and does not explain.}
\noindent The predominant/contextual language introduced in Sec.~\ref{sec:comparative-structure} gives a compact way to describe the $B_0$ behavior of the two-same-one-different family: a \textit{partial-cabling link} realized with probability $\frac{2}{3}$, and \textit{a Borromean-type exception} realized with probability $\frac{1}{3}$. It is worth being explicit that this is a description of the probability-weighted mixture of outcomes already visible in Table~\ref{tab:main-w-summary}, not an
independent derivation of that mixture. As discussed in Sec.~\ref{sec:status-of-correspondence}, none of the link assignments in this paper are obtained from a topological invariant computed independently of the Schmidt-rank calculation; the hybrid classification inherits this same status. Its value is organizational. It lets Table~\ref{tab:topological-summary} express seven states across two bases in a single consistent vocabulary, rather than explanatory in a deeper sense.

\paragraph{Scope of the calculation.}
\noindent Two restrictions of the present analysis are worth flagging directly, beyond those already noted in Sec. \ref{sec:status-of-correspondence}. First, every coefficient matrix $M$ arising in Sections~\ref{sec:ghz-states} and~\ref{sec:w-states} turns out to be diagonal or reducible to a simple $2\times2$ or rank-one-plus-identity form (Sec.~\ref{sec:preliminaries}, Appendices~\ref{app:ghz-measurement}--\ref{app:w-mub-alldifferent}); this is a consequence of measuring in a basis ($B_0$ or an MUB) that is naturally adapted to the permutation symmetry of the states considered, and should not be expected to hold for measurement in a generic, unadapted basis. Second, the states treated here — \textit{permutation-symmetric} $GHZ$ and $W$ \textit{bags} — are only two families among the much larger set of SLOCC-inequivalent entanglement classes available to three qutrits \cite{ec2008}; nothing in this work establishes that the \textit{cabling dictionary}, or even the \textit{qualitative repeated-index/all-different-index distinction} identified above, extends to that larger space.

\section{Conclusion}
\label{sec:conclusion}

\noindent We have extended the Aravind topological correspondence between multipartite entanglement and topological linking \cite{aravind1997, kauffman2002quantum}, and our own earlier qubit-level treatments \cite{sougata2509, sougata2512}, from qubits to qutrits. The extension required replacing single topological rings with \textit{two-strand cables} (Sec.~\ref{sec:qutrit-obstruction}), so that the three possible values of the residual Schmidt rank after a single-particle measurement, $R \in \{1, 2, 3\}$, map onto the number of strands remaining linked, $0$, $1$, or $2$.\\

\noindent Applying this framework to the qutrit $GHZ$ state and to the full family of symmetric qutrit $W$-class states, we found that $|GHZ_3\rangle$ realizes the cabling dictionary in its cleanest possible form: a single well-defined link type per measurement basis, uniform across every outcome (Section~\ref{sec:ghz-states}, Table~\ref{tab:main-ghz-summary}). The \textit{two-same one-different $W$-states} (i.e. $\ket{W_{p,p,q}^{sym}}$) depart from this pattern under computational-basis measurement, requiring the hybrid predominant/contextual classification introduced in Sec.~\ref{sec:comparative-structure}; under MUB measurement they recover a clean, uniform partial-cabling assignment. The \textit{all-different W-state}, $|W_{\{0,1,2\}}\rangle$, was found to behave cleanly in both bases, matching $|GHZ_3\rangle$'s qualitative pattern despite belonging to a structurally different state family (Section~\ref{sec:results-discussion}).\\

\noindent We have been deliberately explicit, in Sec.~\ref{sec:status-of-correspondence} and again above, that the cabling dictionary is a labeling convention built to reproduce an independently-computed Schmidt rank, not a topological invariant derived from the link diagrams themselves. The contribution of this paper is accordingly twofold and should be read as such: (i) a complete, algebraically verified account of measurement-induced entanglement splitting for qutrit $GHZ$- and $W$-class states under both computational and mutually unbiased bases (Appendices~\ref{app:ghz-measurement}--\ref{app:w-mub-alldifferent}),
which stands independently of any topological interpretation; and (ii) a consistent extension of the qubit-level topological vocabulary that organizes those results, whose status as bookkeeping rather than invariant is stated openly rather than implied.\\

\noindent Several directions follow naturally from this work. Deriving the cabling correspondence from an actual topological or algebraic structure — a braid-group representation of the measurement operation, or a polynomial link invariant that changes value in step with $R$ — would be needed to move the dictionary from a descriptive device to a genuine invariant, in the spirit of more general topological link models of multipartite entanglement that have recently been proposed \cite{bao2022topological}, and is in our view the most important open problem raised by this paper. On the algebraic side, extending the present analysis beyond the \textit{permutation-symmetric $GHZ$ and $W$ bags} to the full space of SLOCC-inequivalent three-qutrit classes, and beyond $d=3$ to general qudit dimension $d$ with $(d-1)$-strand cables, would test whether the \textit{repeated-index versus
all-different-index} distinction identified in Section~\ref{sec:results-discussion} is a general feature or an artifact of the qutrit case. Extending the framework to mixed states and to entanglement degradation under noise channels is a further natural continuation. Finally, the measurement-induced rank values computed here are, independently of any topological reading, directly testable predictions for photonic or trapped-ion qutrit platforms capable of preparing $GHZ$ and $W$-class qutrit states and performing MUB measurements, capabilities already demonstrated experimentally for trapped-ion qudit processors \cite{ringbauer2022universal} and for photonic GHZ states beyond the qubit level \cite{ghz32018},and an experimental comparison would be a valuable check of the underlying Schmidt-rank calculation itself.


\appendix
\newpage

\section{Measurement Analysis of Qutrit $GHZ$ States}
\label{app:ghz-measurement}

\subsection{Definition of the State}

\noindent The qutrit $GHZ$ state is the maximally-correlated superposition over all three computational basis (CB) values appearing identically on all three particles:
\begin{equation}
\label{eq:ghz-def}
\ket{GHZ_3} = \frac{1}{\sqrt3}\big(\ket{000}+\ket{111}+\ket{222}\big) = \frac{1}{\sqrt3}\sum_{a=0}^{2}\ket{a}_A\ket{a}_B\ket{a}_C.
\end{equation}
\noindent As with the $W$-states, we measure particle $A$ in the computational basis $B_0=\{\ket0,\ket1,\ket2\}$ using the projectors:
\begin{align}
P_0^A &= \ket{0}_A\bra{0} \otimes I_B \otimes I_C ,\nonumber\\
P_1^A &= \ket{1}_A \bra{1}\otimes I_B \otimes I_C, \nonumber\\
P_2^A &= \ket{2}_A\bra{2} \otimes I_B \otimes I_C,
\end{align}
\noindent and, for the MUB analysis, the same MUB vectors $\ket{e_j}$ and inverse relation $\ket{a}_A=\frac{1}{\sqrt3}\sum_j\varepsilon_{j,a}\ket{e_j}_A$, $\varepsilon_{j,a}=\omega^{-(ja+ma^2)}$, $|\varepsilon_{j,a}|=1$, used throughout the $W$-state appendices.

\subsection{Measurement in the Computational Basis}

\subsubsection{Outcome $\ket{0}_A$}

\begin{equation}
\label{p0ghz}
P_0^A\ket{GHZ_3} = \frac{1}{\sqrt3}\ket{000}_{ABC} = \frac{1}{\sqrt3}\ket{0}_A\otimes\ket{00}_{BC},
\end{equation}
\begin{equation}
p(0) = \frac13, \qquad \ket{\phi_{BC}^{(0)}} = \ket{00}_{BC}.
\end{equation}

\paragraph{Coefficient Matrix and Eigenvalues : }
\begin{equation}
M =
\begin{pmatrix}
1 & 0 & 0 \\
0 & 0 & 0 \\
0 & 0 & 0
\end{pmatrix},
\end{equation}
\noindent and, since $M$ is real and diagonal, $M^\dagger=M$, so
\begin{equation}
M^{\dagger}M =
\begin{pmatrix}
1 & 0 & 0 \\
0 & 0 & 0 \\
0 & 0 & 0
\end{pmatrix}
\begin{pmatrix}
1 & 0 & 0 \\
0 & 0 & 0 \\
0 & 0 & 0
\end{pmatrix}
=
\begin{pmatrix}
1 & 0 & 0 \\
0 & 0 & 0 \\
0 & 0 & 0
\end{pmatrix}.
\end{equation}
\noindent Since this matrix is diagonal, the eigenvalues are simply its diagonal entries:
\begin{equation}
\lambda_1 = 1, \qquad \lambda_2 = 0, \qquad \lambda_3 = 0.
\end{equation}
\noindent One non-zero eigenvalue $\Rightarrow R = 1$ (a product state).

\subsubsection{Outcome $\ket{1}_A$}

\begin{equation}
P_1^A\ket{GHZ_3} = \frac{1}{\sqrt3}\ket{111}_{ABC} = \frac{1}{\sqrt3}\ket{1}_A\otimes\ket{11}_{BC},
\end{equation}
\begin{equation}
p(1) = \frac13, \qquad \ket{\phi_{BC}^{(1)}} = \ket{11}_{BC}.
\end{equation}

\paragraph{Coefficient Matrix and Eigenvalues : }
\begin{equation}
M =
\begin{pmatrix}
0 & 0 & 0 \\
0 & 1 & 0 \\
0 & 0 & 0
\end{pmatrix},
\end{equation}
\noindent and, since $M$ is real and diagonal, $M^\dagger=M$, so
\begin{equation}
M^{\dagger}M =
\begin{pmatrix}
0 & 0 & 0 \\
0 & 1 & 0 \\
0 & 0 & 0
\end{pmatrix}
\begin{pmatrix}
0 & 0 & 0 \\
0 & 1 & 0 \\
0 & 0 & 0
\end{pmatrix}
=
\begin{pmatrix}
0 & 0 & 0 \\
0 & 1 & 0 \\
0 & 0 & 0
\end{pmatrix}.
\end{equation}
\noindent Since this matrix is diagonal, the eigenvalues are simply its diagonal entries:
\begin{equation}
\lambda_1 = 0, \qquad \lambda_2 = 1, \qquad \lambda_3 = 0.
\end{equation}
\noindent One non-zero eigenvalue $\Rightarrow R = 1$ (a product state).

\subsubsection{Outcome $\ket{2}_A$}

\begin{equation}
P_2^A\ket{GHZ_3} = \frac{1}{\sqrt3}\ket{222}_{ABC} = \frac{1}{\sqrt3}\ket{2}_A\otimes\ket{22}_{BC},
\end{equation}
\begin{equation}
p(2) = \frac13, \qquad \ket{\phi_{BC}^{(2)}} = \ket{22}_{BC}.
\end{equation}

\paragraph{Coefficient Matrix and Eigenvalues : }
\begin{equation}
M =
\begin{pmatrix}
0 & 0 & 0 \\
0 & 0 & 0 \\
0 & 0 & 1
\end{pmatrix},
\end{equation}
\noindent and, since $M$ is real and diagonal, $M^\dagger=M$, so
\begin{equation}
M^{\dagger}M =
\begin{pmatrix}
0 & 0 & 0 \\
0 & 0 & 0 \\
0 & 0 & 1
\end{pmatrix}
\begin{pmatrix}
0 & 0 & 0 \\
0 & 0 & 0 \\
0 & 0 & 1
\end{pmatrix}
=
\begin{pmatrix}
0 & 0 & 0 \\
0 & 0 & 0 \\
0 & 0 & 1
\end{pmatrix}.
\end{equation}
\noindent Since this matrix is diagonal, the eigenvalues are simply its diagonal entries:
\begin{equation}
\lambda_1 = 0, \qquad \lambda_2 = 0, \qquad \lambda_3 = 1.
\end{equation}
\noindent One non-zero eigenvalue $\Rightarrow R = 1$ (a product state).\\

\noindent Every computational-basis outcome therefore leaves $B$ and $C$ in an unentangled product state the correlations in $\ket{GHZ_3}$ are entirely classical (diagonal) in this basis.

\subsection{Measurement in a Mutually Unbiased Basis}

\noindent Apply the inverse relation to the $A$ register in \eqref{eq:ghz-def} and collect the coefficient of each $\ket{e_j}_A$:
\begin{align}
\ket{GHZ_3} &= \frac{1}{\sqrt3}\sum_{a=0}^{2}\left(\frac{1}{\sqrt3}\sum_j\varepsilon_{j,a}\ket{e_j}_A\right)\otimes\ket{aa}_{BC}\nonumber\\
&= \frac13\sum_{j=0}^{2}\ket{e_j}_A\otimes\ket{\chi_j}_{BC}, \qquad \ket{\chi_j}_{BC} = \sum_{a=0}^{2}\varepsilon_{j,a}\ket{aa}_{BC}.
\end{align}
\noindent The three kets $\ket{00},\ket{11},\ket{22}$ are mutually orthogonal and every $|\varepsilon_{j,a}|=1$, so by inspection:
\begin{equation}
\|\chi_j\|^2 = 3, \qquad \text{for every } j,m,
\end{equation}
\noindent giving measurement probabilities
\begin{equation}
p(j) = \frac19\|\chi_j\|^2 = \frac13, \qquad p(0)=p(1)=p(2)=\frac13.
\end{equation}
\noindent Dividing by $\|\chi_j\|=\sqrt3$, the normalized post-measurement state is:
\begin{equation}
\ket{\phi_j}_{BC} = \frac{1}{\sqrt3}\Big[\varepsilon_{j,0}\ket{00}+\varepsilon_{j,1}\ket{11}+\varepsilon_{j,2}\ket{22}\Big].
\end{equation}

\paragraph{Coefficient Matrix and Eigenvalues : }
 \noindent Reading off the coefficient of each $\ket{b}_B\ket{c}_C$ term, ordered as $(\ket0,\ket1,\ket2)$ on both $B$ and $C$:
\begin{equation}
M_j = \frac{1}{\sqrt3}\begin{pmatrix}\varepsilon_{j,0} & 0 & 0\\ 0 & \varepsilon_{j,1} & 0\\ 0 & 0 & \varepsilon_{j,2}\end{pmatrix},
\end{equation}
\noindent and
\begin{equation}
M_j^\dagger = \frac{1}{\sqrt3}\begin{pmatrix}\overline{\varepsilon_{j,0}} & 0 & 0\\ 0 & \overline{\varepsilon_{j,1}} & 0\\ 0 & 0 & \overline{\varepsilon_{j,2}}\end{pmatrix}.
\end{equation}
\noindent Thus
\begin{equation}
M_j^\dagger M_j = \frac13\begin{pmatrix}|\varepsilon_{j,0}|^2 & 0 & 0\\ 0 & |\varepsilon_{j,1}|^2 & 0\\ 0 & 0 & |\varepsilon_{j,2}|^2\end{pmatrix} = \frac13\begin{pmatrix}1 & 0 & 0\\ 0 & 1 & 0\\ 0 & 0 & 1\end{pmatrix},
\end{equation}
\noindent using $|\varepsilon_{j,a}|=1$ for every $j,a,m$. This matrix is already diagonal, so the eigenvalues are simply its diagonal entries:
\begin{equation}
\lambda_1 = \lambda_2 = \lambda_3 = \frac13
\end{equation}
\noindent All three eigenvalues are strictly positive and equal $\Rightarrow R=3$.\\

\noindent The result is independent of the specific phases $\varepsilon_{j,a}$ beyond their modulus, and therefore holds for every outcome $j$ and every MUB $m$.

\subsection{Summary Table}

\begin{table}[H]
\centering
\renewcommand{\arraystretch}{1.5}
\begin{tabular}{c|c|c|c|c|c}
\toprule
Basis measured on $A$ & Outcome $j$ & $\ket{\phi_{BC}}$ & $p(j)$ & Eigenvalues $\{\lambda_i\}$ & $R$ \\
\midrule
$B_0$ (computational) & $0,1,2$ & $\ket{jj}$ & $\frac{1}{3}$ & $\{1,0,0\}$ & 1 \\
\midrule
Any MUB ($B_1,B_2,B_3$; $m=0,1,2$) & $0,1,2$ & $\frac{1}{\sqrt3}\sum_a\varepsilon_{j,a}\ket{aa}$ & $\frac{1}{3}$ & $\left\{\tfrac13,\tfrac13,\tfrac13\right\}$ & 3 \\
\bottomrule
\end{tabular}
\caption{Summary of measurement outcomes, probabilities, post-measurement states, eigenvalues of $M^{\dagger}M$, and Schmidt ranks for $\ket{GHZ_3}$, measured in the computational basis and in a Mutually Unbiased Basis.}
\end{table}


\section{Measurement Analysis of the Two-Same-One-Different Symmetric Qutrit W-States in the Computational Basis}
\label{app:wstates}

\subsection{Definition of the States}

\noindent To systematically analyze the topological entanglement structure of $W$-class states in a qutrit system ($d=3$), we define the complete family of symmetric non-trivial qutrit $W$-states. These states are constructed by taking uniform superpositions of all distinct permutations of a specific \textit{bag} (multiset) of computational basis indices.\\

\noindent For the \textit{two-same one-different} index distributions, fixing two distinct symbols $p,q\in\{0,1,2\}$, the general member of the family is:
\begin{equation}
\label{eq:Wppq-def-compbasis}
\ket{W_{p,p,q}^{sym}} = \frac{1}{\sqrt3}\big(\ket{ppq}+\ket{pqp}+\ket{qpp}\big).
\end{equation}
\noindent There are exactly six ordered pairs $(p,q)$ with $p\neq q$, giving the six fundamental states:
\begin{align}
\ket{W_1^{sym}} &= \ket{W_{0,0,1}^{sym}} = \frac{1}{\sqrt{3}} (\ket{001} + \ket{010} + \ket{100}) \nonumber\\
\ket{W_2^{sym}} &= \ket{W_{0,0,2}^{sym}} = \frac{1}{\sqrt{3}} (\ket{002} + \ket{020} + \ket{200}) \nonumber\\
\ket{W_3^{sym}} &= \ket{W_{1,1,0}^{sym}} = \frac{1}{\sqrt{3}} (\ket{110} + \ket{101} + \ket{011}) \nonumber\\
\ket{W_4^{sym}} &= \ket{W_{1,1,2}^{sym}} = \frac{1}{\sqrt{3}} (\ket{112} + \ket{121} + \ket{211}) \nonumber\\
\ket{W_5^{sym}} &= \ket{W_{2,2,0}^{sym}} = \frac{1}{\sqrt{3}} (\ket{220} + \ket{202} + \ket{022}) \nonumber\\
\ket{W_6^{sym}} &= \ket{W_{2,2,1}^{sym}} = \frac{1}{\sqrt{3}} (\ket{221} + \ket{212} + \ket{122})
\end{align}

\noindent Because these states exhibit perfect permutation symmetry, a measurement on any single particle (A, B, or C) yields mathematically equivalent results for the residual bipartite system. Without loss of generality, we perform all projective measurements on qubit A in the computational basis $B_0=\{\ket{0},\ket{1},\ket{2}\}$ using the projectors:

\begin{align}
P_0^A &= \ket{0}_A\bra{0} \otimes I_B \otimes I_C, \nonumber\\
P_1^A &= \ket{1}_A\bra{1} \otimes I_B \otimes I_C, \nonumber\\
P_2^A &= \ket{2}_A\bra{2} \otimes I_B \otimes I_C.
\end{align}

\subsection{General Measurement of $\ket{W_{p,p,q}^{sym}}$}

\noindent Rather than repeat this calculation six times, one for each bag, we exploit the fact that \eqref{eq:Wppq-def-compbasis} is already sorted by the value carried on particle $A$. Let $r$ denote the third symbol of $\{0,1,2\}$, distinct from both $p$ and $q$ ($r\neq p,q$), so that $\{p,q,r\}=\{0,1,2\}$. Since $A$ is measured directly in the computational basis in which \eqref{eq:Wppq-def-compbasis} is already written, the projectors $P_p^A$, $P_q^A$, $P_r^A$ act term-by-term with no basis change required — in contrast to the MUB case of Appendix~\ref{app:mub-wstates}, where a change of basis on $A$ was the central step.

\subsubsection{Outcome $\ket{p}_A$}

\noindent Applying $P_p^A$ to \eqref{eq:Wppq-def-compbasis} selects the two terms carrying $A=p$:
\begin{equation}
P_p^A \ket{W_{p,p,q}^{sym}} = \frac{1}{\sqrt{3}} \ket{p}_A \otimes (\ket{pq} + \ket{qp})_{BC}.
\end{equation}
\noindent Because $p\neq q$, the kets $\ket{pq}_{BC}$ and $\ket{qp}_{BC}$ are orthonormal, so the probability of this outcome is
\begin{equation}
p(p) = \left\Vert \frac{1}{\sqrt{3}} (\ket{pq} + \ket{qp})_{BC} \right\Vert^2 = \frac23,
\end{equation}
\noindent independent of the specific values of $p$ and $q$, and the normalized post-measurement state is
\begin{equation}
\ket{\phi_{BC}^{(p)}} = \frac{1}{\sqrt{2}} (\ket{pq} + \ket{qp})_{BC}.
\end{equation}

\paragraph{Coefficient Matrix and Eigenvalues : } 
\noindent Restricting to the coefficients of $\ket{p}_B\ket{q}_C$ and $\ket{q}_B\ket{p}_C$ (all other entries of the $3\times 3$ coefficient matrix, including every entry involving $r$, are zero), $M^\dagger M$ is diagonal with a single nonzero pair of entries:
\begin{equation}
(M^\dagger M)_{pp} = \tfrac12, \qquad (M^\dagger M)_{qq} = \tfrac12, \qquad (M^\dagger M)_{rr} = 0
\end{equation}
\noindent so that
\begin{equation}
\lambda_p = \tfrac12, \qquad \lambda_q = \tfrac12, \qquad \lambda_r = 0.
\end{equation}
\noindent Two non-zero eigenvalues $\Rightarrow R = 2$, for every choice of $(p,q)$.

\subsubsection{Outcome $\ket{q}_A$}

\noindent Applying $P_q^A$ selects the single remaining term carrying $A=q$:
\begin{equation}
P_q^A \ket{W_{p,p,q}^{sym}} = \frac{1}{\sqrt{3}} \ket{q}_A \otimes \ket{pp}_{BC}
\end{equation}
\noindent so that $p(q) = \frac{1}{3}$ and the normalized state is $\ket{\phi_{BC}^{(q)}} = \ket{pp}_{BC}$. The coefficient matrix has a single nonzero entry, $(M)_{pp}=1$, so $M^\dagger M$ has a single nonzero eigenvalue:
\begin{equation}
\lambda_p = 1, \qquad \lambda_q = 0, \qquad \lambda_r = 0.
\end{equation}
\noindent One non-zero eigenvalue $\Rightarrow R = 1$, for every choice of $(p,q)$.

\subsubsection{Outcome $\ket{r}_A$}

\noindent The symbol $r$ never appears in any of the three terms of \eqref{eq:Wppq-def-compbasis}, so
\begin{equation}
P_r^A \ket{W_{p,p,q}^{sym}} = 0
\end{equation}
\noindent and $p(r) = 0$; this outcome never occurs, for every choice of $(p,q)$.

\subsection{Instantiating the Six States}

\noindent The general result above holds for \emph{every} ordered pair $(p,q)$ with $p\neq q$, and therefore applies uniformly to all \textit{six bags} of Table~\ref{tab:pq-instantiation}: the outcome $\ket{p}_A$ always occurs with probability $\frac{2}{3}$ and leaves $R=2$, the outcome $\ket{q}_A$ always occurs with probability $\frac{1}{3}$ and leaves $R=1$, and $\ket{r}_A$ never occurs. The full outcome-by-outcome results for all six states, obtained by substituting each $(p,q,r)$ triple into this template, are collected in Table~\ref{tab:w2w} of Appendix~\ref{app:comp-summary}.

\begin{table}[H]
\centering
\caption{The six \textit{two-same one-different bags} as instances of $(p,q,r)$.}
\label{tab:pq-instantiation}
\begin{tabular}{c|c|c}
\toprule
State & $(p,q)$ & $r$ \\
\midrule
$\ket{W_1^{sym}}$ & $(0,1)$ & $2$ \\
$\ket{W_2^{sym}}$ & $(0,2)$ & $1$ \\
$\ket{W_3^{sym}}$ & $(1,0)$ & $2$ \\
$\ket{W_4^{sym}}$ & $(1,2)$ & $0$ \\
$\ket{W_5^{sym}}$ & $(2,0)$ & $1$ \\
$\ket{W_6^{sym}}$ & $(2,1)$ & $0$ \\
\bottomrule
\end{tabular}
\end{table}


\section{Measurement Analysis of the All-Different Qutrit $W$-State in the Computational Basis}
\label{app:w012}

\noindent We now repeat the computational-basis analysis of Appendix~\ref{app:wstates} for the remaining state, which fully utilizes the three-dimensional nature of the qutrit space -- the all-different fully symmetric qutrit state:

\begin{equation}
\ket{W_{\{0,1,2\}}} = \frac{1}{\sqrt{6}} (\ket{012} + \ket{021} + \ket{102} + \ket{120} + \ket{201} + \ket{210}).
\end{equation}

\noindent As in Appendix~\ref{app:wstates}, we measure particle $A$ in the computational basis using the projectors $P_0^A,P_1^A,P_2^A$ defined there, and apply the same coefficient-matrix method of Section~\ref{sec:schmidt-decomp}.

\subsection{Measurement of $\ket{W_{\{0,1,2\}}}$}

\subsubsection{Outcome $\ket{0}_A$}

\begin{align}
P_0^A \ket{W_{\{0,1,2\}}} &= \frac{1}{\sqrt{6}} (\ket{012} + \ket{021}) \nonumber\\
&= \frac{1}{\sqrt{6}} \ket{0}_A \otimes (\ket{12} + \ket{21})_{BC}.
\end{align}
\begin{equation}
p(0) = \frac13, \qquad \ket{\phi_{BC}^{(0)}} = \frac{1}{\sqrt{2}} (\ket{12} + \ket{21})_{BC}
\end{equation}

\paragraph{Coefficient Matrix and Eigenvalues : }
\begin{equation}
M =
\begin{pmatrix}
0 & 0 & 0 \\
0 & 0 & \frac{1}{\sqrt{2}} \\
0 & \frac{1}{\sqrt{2}} & 0
\end{pmatrix}
\end{equation}
\begin{equation}
M^{\dagger}M =
\begin{pmatrix}
0 & 0 & 0 \\
0 & \tfrac12 & 0 \\
0 & 0 & \tfrac12
\end{pmatrix}.
\end{equation}
\noindent Since this matrix is diagonal, the eigenvalues are simply its diagonal entries:
\begin{equation}
\lambda_1 = 0, \qquad \lambda_2 = \tfrac12, \qquad \lambda_3 = \tfrac12
\end{equation}
\noindent Two non-zero eigenvalues $\Rightarrow R = 2$.

\subsubsection{Outcome $\ket{1}_A$}

\begin{align}
P_1^A \ket{W_{\{0,1,2\}}} &= \frac{1}{\sqrt{6}} (\ket{102} + \ket{120}) \nonumber\\
&= \frac{1}{\sqrt{6}} \ket{1}_A \otimes (\ket{02} + \ket{20})_{BC}
\end{align}
\begin{equation}
p(1) = \frac13, \qquad \ket{\phi_{BC}^{(1)}} = \frac{1}{\sqrt{2}} (\ket{02} + \ket{20})_{BC}
\end{equation}

\paragraph{Coefficient Matrix and Eigenvalues : }
\begin{equation}
M =
\begin{pmatrix}
0 & 0 & \frac{1}{\sqrt{2}} \\
0 & 0 & 0 \\
\frac{1}{\sqrt{2}} & 0 & 0
\end{pmatrix}
\end{equation}
\begin{equation}
M^{\dagger}M =
\begin{pmatrix}
\tfrac12 & 0 & 0 \\
0 & 0 & 0 \\
0 & 0 & \tfrac12
\end{pmatrix}.
\end{equation}
\noindent Since this matrix is diagonal, the eigenvalues are simply its diagonal entries:
\begin{equation}
\lambda_1 = \tfrac12, \qquad \lambda_2 = 0, \qquad \lambda_3 = \tfrac12
\end{equation}
\noindent Two non-zero eigenvalues $\Rightarrow R = 2$.

\subsubsection{Outcome $\ket{2}_A$}

\begin{align}
P_2^A \ket{W_{\{0,1,2\}}} &= \frac{1}{\sqrt{6}} (\ket{201} + \ket{210}) \nonumber\\
&= \frac{1}{\sqrt{6}} \ket{2}_A \otimes (\ket{01} + \ket{10})_{BC}
\end{align}
\begin{equation}
p(2) = \frac13, \qquad \ket{\phi_{BC}^{(2)}} = \frac{1}{\sqrt{2}} (\ket{01} + \ket{10})_{BC}
\end{equation}

\paragraph{Coefficient Matrix and Eigenvalues : }
\begin{equation}
M =
\begin{pmatrix}
0 & \frac{1}{\sqrt{2}} & 0 \\
\frac{1}{\sqrt{2}} & 0 & 0 \\
0 & 0 & 0
\end{pmatrix}
\end{equation}
\begin{equation}
M^{\dagger}M=
\begin{pmatrix}
\tfrac12 & 0 & 0 \\
0 & \tfrac12 & 0 \\
0 & 0 & 0
\end{pmatrix}.
\end{equation}
\noindent Since this matrix is diagonal, the eigenvalues are simply its diagonal entries:
\begin{equation}
\lambda_1 = \tfrac12, \qquad \lambda_2 = \tfrac12, \qquad \lambda_3 = 0,
\end{equation}
\noindent Two non-zero eigenvalues $\Rightarrow R = 2$.


\section{Summary of Computational-Basis Measurement Results}
\label{app:comp-summary}

\noindent The following table summarizes the measurement outcomes for all symmetric qutrit $W$-states, both the \textit{two same one-different} states of Appendix~\ref{app:wstates} and the \textit{all-different} state of Appendix~\ref{app:w012}, when particle $A$ is measured in the standard computational basis.

\begin{table}[H]
\centering
\caption{Summary of measurement outcomes, probabilities, post-measurement states, eigenvalues of $M^{\dagger}M$ obtained from the characteristic equation, and Schmidt ranks for all symmetric qutrit $W$-states}
\label{tab:w2w}
\begin{tabular}{c|c|c|c|c|c}
\toprule
State & Outcome on $A$ & $\ket{\phi_{BC}}$ & Probability & Eigenvalues $\{\lambda_i\}$ & $R$ \\
\midrule
\multirow{3}{*}{$W_1$} & $\ket{0}$ & $\frac{1}{\sqrt{2}}(\ket{01}+\ket{10})$ & $2/3$ & $\{\tfrac12,\tfrac12,0\}$ & 2 \\
                        & $\ket{1}$ & $\ket{00}$ & $1/3$ & $\{1,0,0\}$ & 1 \\
                        & $\ket{2}$ & --- & $0$ & --- & --- \\
\midrule
\multirow{3}{*}{$W_2$} & $\ket{0}$ & $\frac{1}{\sqrt{2}}(\ket{02}+\ket{20})$ & $2/3$ & $\{\tfrac12,0,\tfrac12\}$ & 2 \\
                        & $\ket{1}$ & --- & $0$ & --- & --- \\
                        & $\ket{2}$ & $\ket{00}$ & $1/3$ & $\{1,0,0\}$ & 1 \\
\midrule
\multirow{3}{*}{$W_3$} & $\ket{0}$ & $\ket{11}$ & $1/3$ & $\{0,1,0\}$ & 1 \\
                        & $\ket{1}$ & $\frac{1}{\sqrt{2}}(\ket{01}+\ket{10})$ & $2/3$ & $\{\tfrac12,\tfrac12,0\}$ & 2 \\
                        & $\ket{2}$ & --- & $0$ & --- & --- \\
\midrule
\multirow{3}{*}{$W_4$} & $\ket{0}$ & --- & $0$ & --- & --- \\
                        & $\ket{1}$ & $\frac{1}{\sqrt{2}}(\ket{12}+\ket{21})$ & $2/3$ & $\{0,\tfrac12,\tfrac12\}$ & 2 \\
                        & $\ket{2}$ & $\ket{11}$ & $1/3$ & $\{0,1,0\}$ & 1 \\
\midrule
\multirow{3}{*}{$W_5$} & $\ket{0}$ & $\ket{22}$ & $1/3$ & $\{0,0,1\}$ & 1 \\
                        & $\ket{1}$ & --- & $0$ & --- & --- \\
                        & $\ket{2}$ & $\frac{1}{\sqrt{2}}(\ket{02}+\ket{20})$ & $2/3$ & $\{\tfrac12,0,\tfrac12\}$ & 2 \\
\midrule
\multirow{3}{*}{$W_6$} & $\ket{0}$ & --- & $0$ & --- & --- \\
                        & $\ket{1}$ & $\ket{22}$ & $1/3$ & $\{0,0,1\}$ & 1 \\
                        & $\ket{2}$ & $\frac{1}{\sqrt{2}}(\ket{12}+\ket{21})$ & $2/3$ & $\{0,\tfrac12,\tfrac12\}$ & 2 \\
\midrule
\multirow{3}{*}{$W_{\{0,1,2\}}$} & $\ket{0}$ & $\frac{1}{\sqrt{2}}(\ket{12}+\ket{21})$ & $1/3$ & $\{0,\tfrac12,\tfrac12\}$ & 2 \\
                                  & $\ket{1}$ & $\frac{1}{\sqrt{2}}(\ket{02}+\ket{20})$ & $1/3$ & $\{\tfrac12,0,\tfrac12\}$ & 2 \\
                                  & $\ket{2}$ & $\frac{1}{\sqrt{2}}(\ket{01}+\ket{10})$ & $1/3$ & $\{\tfrac12,\tfrac12,0\}$ & 2 \\
\bottomrule
\end{tabular}
\end{table}


\section{Measurement of the Two-Same-One-Different $W$-States in a Mutually Unbiased Basis}
\label{app:mub-wstates}

\noindent Appendix~\ref{app:wstates} measured particle $A$ in the computational basis. We now repeat the analysis measuring $A$ in one of the three non-computational mutually unbiased bases (MUBs) instead, and show that the result is the same for all six states $\ket{W_1^{sym}},\dots,\ket{W_6^{sym}}$, all three measurement outcomes, and all three MUBs simultaneously.\\

\noindent This section treats the \textit{two-same one-different} family; we refer to it as Part A.

\subsection{Setup}

\noindent The three non-computational MUBs for a qutrit are indexed by $m\in\{0,1,2\}$, with basis vectors
\begin{equation}
\ket{e_j} = \frac{1}{\sqrt3}\sum_{k=0}^{2}\omega^{jk+mk^2}\ket{k}, \qquad j=0,1,2, \qquad \omega = e^{2\pi i/3}.
\end{equation}
\noindent A full construction and derivation of this basis is given in the Preliminaries. The one fact we need here is the inverse relation, expressing a computational basis vector $\ket{a}$ in terms of the $\ket{e_j}$'s. It follows from completeness of $\{\ket{e_j}\}$ together with $\langle e_j\vert a\rangle =\overline{\langle a\vert e_j\rangle}$, and reads:
\begin{equation}
\label{eq:inverse-mub}
\ket{a} = \frac{1}{\sqrt3}\sum_{j=0}^{2}\omega^{-(ja+ma^2)}\ket{e_j}, \qquad a\in\{0,1,2\}.
\end{equation}

\subsection{The State, Grouped by the Value on Particle A}

\noindent Fix two distinct symbols $p,q\in\{0,1,2\}$. The symmetric state with $p$ appearing twice and $q$ once is:
\begin{equation}
\ket{W_{p,p,q}^{sym}} = \frac{1}{\sqrt3}\big(\ket{ppq}+\ket{pqp}+\ket{qpp}\big).
\end{equation}
\noindent Each of the six states from Appendix~\ref{app:wstates} is one instance of this for a specific $(p,q)$: for example $\ket{W_1^{sym}}$ is $(p,q)=(0,1)$, and $\ket{W_4^{sym}}$ is $(p,q)=(1,2)$.\\

\noindent We sort the three terms by what particle $A$ carries. The first two terms both have $A=p$, and the third has $A=q$:
\begin{itemize}
\item $\ket{ppq} = \ket{p}_A\ket{pq}_{BC}$.
\item $\ket{pqp} = \ket{p}_A\ket{qp}_{BC}$.
\item $\ket{qpp} = \ket{q}_A\ket{pp}_{BC}$.
\end{itemize}
\noindent Collecting the two terms that share $\ket{p}_A$ gives:
\begin{equation}
\label{eq:Wppq_grouped}
\ket{W_{p,p,q}^{sym}} = \frac{1}{\sqrt3}\Big[\ket{p}_A\big(\ket{pq}_{BC}+\ket{qp}_{BC}\big) + \ket{q}_A\ket{pp}_{BC}\Big].
\end{equation}

\subsection{Rewriting $\ket{p}_A$ and $\ket{q}_A$ in the MUB Basis}

\noindent Apply the inverse relation \eqref{eq:inverse-mub} once with $a=p$ and once with $a=q$. To keep the expressions short, define:
\begin{equation}
\gamma_j = \omega^{-(jp+mp^2)}, \qquad \delta_j = \omega^{-(jq+mq^2)},
\end{equation}
\noindent so that:
\begin{equation}
\ket{p}_A = \frac{1}{\sqrt3}\sum_{j=0}^2 \gamma_j\,\ket{e_j}_A, \qquad \ket{q}_A = \frac{1}{\sqrt3}\sum_{j=0}^2 \delta_j\,\ket{e_j}_A
\end{equation}

\noindent Since $\omega=e^{2\pi i/3}$ lies on the unit circle, every integer power of $\omega$ also has modulus $1$. Both $\gamma_j$ and $\delta_j$ are such powers, so:
\begin{equation}
\label{eq:unitmod}
|\gamma_j| = 1, \qquad |\delta_j| = 1, \qquad \text{for every } j\in\{0,1,2\} \text{ and every } m\in\{0,1,2\}
\end{equation}
\noindent Every result in this section follows from this one fact, and from nothing more specific about $j$, $m$, $p$, or $q$.

\subsection{Assembling the State in the MUB Basis}

\noindent Substitute both expansions into \eqref{eq:Wppq_grouped}:
\begin{align}
\ket{W_{p,p,q}^{sym}} &= \frac{1}{\sqrt3}\left[\left(\frac{1}{\sqrt3}\sum_j \gamma_j\ket{e_j}_A\right)\big(\ket{pq}+\ket{qp}\big)_{BC} + \left(\frac{1}{\sqrt3}\sum_j \delta_j\ket{e_j}_A\right)\ket{pp}_{BC}\right]\nonumber\\
&= \frac13\sum_{j=0}^2 \ket{e_j}_A \otimes \ket{\chi_j}_{BC},
\end{align}
\noindent where the two factors of $\frac{1}{\sqrt3}$ combine to give the overall $\frac13$, and where we collected everything multiplying $\ket{e_j}_A$ into a single ket on $BC$:
\begin{equation}
\label{eq:chij_A}
\ket{\chi_j}_{BC} = \gamma_j\big(\ket{pq}+\ket{qp}\big) + \delta_j\ket{pp}.
\end{equation}
\noindent Note that $\ket{\chi_j}$ does \emph{not} include the outer $\frac13$; that factor stays in front of the sum over $j$.

\subsection{Norm of the Unnormalized Branch $\ket{\chi_j}$}

\noindent The three kets $\ket{pq}$, $\ket{qp}$, $\ket{pp}$ are distinct computational basis vectors of $B\otimes C$ (because $p\ne q$), so they are mutually orthogonal. For a vector expanded in an orthonormal set, the norm-squared is just the sum of the squared moduli of its coefficients:
\begin{equation}
\|\chi_j\|^2 = |\gamma_j|^2 + |\gamma_j|^2 + |\delta_j|^2.
\end{equation}
\noindent Using $|\gamma_j|=|\delta_j|=1$ from \eqref{eq:unitmod}:
\begin{equation}
\label{eq:normA}
\|\chi_j\|^2 = 1+1+1 = 3.
\end{equation}
\noindent This value does not depend on $j$, $m$, $p$, or $q$ — only on the fact that there are three terms, each of modulus $1$.

\subsection{Measurement Probabilities}

\noindent The unnormalized post-measurement branch for outcome $j$ is $P_j^A\ket{W_{p,p,q}^{sym}} = \frac13\ket{e_j}_A\otimes\ket{\chi_j}_{BC}$. Its probability is the squared norm:
\begin{equation}
\label{eq:probA}
p(j) = \left\|\frac13\ket{\chi_j}\right\|^2 = \left(\frac13\right)^2\|\chi_j\|^2 = \frac19\cdot3 = \frac13.
\end{equation}
\noindent Since this holds for every $j$, all three outcomes are equally likely:
\begin{equation}
p(0)=p(1)=p(2)=\frac13.
\end{equation}

\subsection{The Normalized Post-Measurement State}

\noindent Dividing $\ket{\chi_j}$ by its own norm $\|\chi_j\|=\sqrt3$ gives the normalized residual state:
\begin{equation}
\ket{\phi_j}_{BC} = \frac{1}{\sqrt3}\Big[\gamma_j\big(\ket{pq}+\ket{qp}\big)+\delta_j\ket{pp}\Big].
\end{equation}
\noindent A global phase does not affect a state's Schmidt decomposition, so we may factor out $\gamma_j$ and drop it.\\

\noindent Define $r_j = \delta_j/\gamma_j$; since $\gamma_j$ and $\delta_j$ both have modulus $1$, so does their ratio, $|r_j|=1$. Up to this irrelevant global phase:
\begin{equation}
\label{eq:phij_final}
\ket{\phi_j}_{BC} \;\propto\; \frac{1}{\sqrt3}\Big[\big(\ket{pq}+\ket{qp}\big) + r_j\ket{pp}\Big].
\end{equation}

\subsection{Schmidt Rank via the Coefficient Matrix}

\noindent Only the levels $p$ and $q$ appear in \eqref{eq:phij_final}, so we restrict to the two-dimensional subspace spanned by $\{\ket{p},\ket{q}\}$ on each of $B$ and $C$, ordered as $(\ket{p},\ket{q})$. Reading off the coefficient of each basis term:
\begin{itemize}
\item coefficient of $\ket{p}_B\ket{p}_C$ is $\dfrac{r_j}{\sqrt3}$.
\item coefficient of $\ket{p}_B\ket{q}_C$ is $\dfrac{1}{\sqrt3}$.
\item coefficient of $\ket{q}_B\ket{p}_C$ is $\dfrac{1}{\sqrt3}$.
\item coefficient of $\ket{q}_B\ket{q}_C$ is $0$
\end{itemize}
\noindent the coefficient matrix (rows indexed by $B\in\{p,q\}$, columns by $C\in\{p,q\}$) is:
\begin{equation}
M_j = \frac{1}{\sqrt3}\begin{pmatrix} r_j & 1\\ 1 & 0\end{pmatrix}.
\end{equation}

\paragraph{Conjugate Transpose : }
\noindent $M_j$ has one complex entry, $r_j$; the rest are real. Taking the conjugate transpose swaps rows and columns and conjugates each entry:
\begin{equation}
M_j^\dagger = \frac{1}{\sqrt3}\begin{pmatrix}\overline{r_j} & 1\\ 1 & 0\end{pmatrix}
\end{equation}

\paragraph{$M_j^\dagger M_j$ :}
\begin{align}
(M_j^\dagger M_j)_{11} &= \frac13\big(\overline{r_j}\cdot r_j + 1\cdot 1\big) = \frac13\big(|r_j|^2+1\big) = \frac13(1+1)=\frac23.\nonumber\\
(M_j^\dagger M_j)_{12} &= \frac13\big(\overline{r_j}\cdot 1 + 1\cdot 0\big) = \frac{\overline{r_j}}{3}.\nonumber\\
(M_j^\dagger M_j)_{21} &= \frac13\big(1\cdot r_j + 0\cdot 1\big) = \frac{r_j}{3}.\nonumber\\
(M_j^\dagger M_j)_{22} &= \frac13\big(1\cdot 1 + 0\cdot 0\big) = \frac13.
\end{align}
so that:
\begin{equation}
M_j^\dagger M_j = \begin{pmatrix}\dfrac23 & \dfrac{\overline{r_j}}{3}\\[6pt] \dfrac{r_j}{3} & \dfrac13\end{pmatrix}
\end{equation}
\noindent This matrix is Hermitian, as it must be: the two off-diagonal entries are complex conjugates of each other. Notice also that the diagonal entries, $\frac23$ and $\frac13$, do not depend on $r_j$ at all; only the off-diagonal phase does.

\subsection{Eigenvalues via the Characteristic Equation}

\noindent For a $2\times2$ Hermitian matrix, the characteristic equation takes the compact form $\lambda^2 - (\mathrm{Tr})\lambda + \det = 0$, where $\mathrm{Tr}$ and $\det$ are the trace and determinant of the matrix.

\paragraph{Trace : }
\begin{equation}
\mathrm{Tr}\big(M_j^\dagger M_j\big) = \frac23+\frac13 = 1
\end{equation}
\noindent This must equal $1$: since $M_j^\dagger M_j$ is a properly normalized reduced density matrix, its trace is the total probability, which is always $1$.

\paragraph{Determinant : }
\begin{equation}
\det\big(M_j^\dagger M_j\big) = \frac23\cdot\frac13 - \frac{\overline{r_j}}{3}\cdot\frac{r_j}{3} = \frac29 - \frac{|r_j|^2}{9} = \frac29-\frac19 = \frac19.
\end{equation}
\noindent The step $|r_j|^2=1$ is exactly where the unbiasedness property enters the final numerical answer.

\paragraph{Characteristic Equation and Its Roots :}
\begin{equation}
\lambda^2 - \lambda + \frac19 = 0.
\end{equation}
\noindent Solving with the quadratic formula:
\begin{equation}
\lambda_\pm = \frac{1\pm\sqrt{1-4\cdot\frac19}}{2} = \frac{1\pm\sqrt{\frac59}}{2} = \frac{1\pm\frac{\sqrt5}{3}}{2} = \frac{3\pm\sqrt5}{6}.
\end{equation}
\noindent Both roots are strictly positive (since $\sqrt5<3$), so both are non-zero eigenvalues, giving Schmidt rank $R=2$.

\subsection{Result}

\begin{equation}
\boxed{p(j)=\frac13,\quad R=2, \quad \lambda_\pm=\frac{3\pm\sqrt5}{6} \quad \text{for every } j\in\{0,1,2\}, m\in\{0,1,2\}, (p,q)}.
\end{equation}
\noindent The entire derivation used only $|\gamma_j|=|\delta_j|=1$, the defining unbiasedness property of the MUB vectors, and never used the specific numerical values of $j$, $m$, $p$, or $q$. This is why the same three numbers, $p(j)=\frac13$, $R=2$, and $\lambda_\pm=\frac{3\pm\sqrt5}{6}$, appear for all six states $\ket{W_1^{sym}},\dots,\ket{W_6^{sym}}$, all three measurement outcomes, and all three MUBs at once: the uniformity is a structural consequence of unbiasedness, not eighteen separate coincidences.


\section{Measurement of the All-Different W-State in a Mutually Unbiased Basis}
\label{app:mub-w012}\label{app:w-mub-alldifferent}

\noindent In this section, we analyze the all-different state, $\ket{W_{\{0,1,2\}}}$. As shown in Appendix~\ref{app:w012}, measuring particle $A$ in the standard computational basis leaves the residual pair ($B$ and $ C $) with a Schmidt rank of $R=2$. Here, we will demonstrate that switching our measurement on A to a Mutually Unbiased Basis (MUB) acts as a phase scrambler, forcing the remaining $ B-C $ pair into a state with the maximum possible Schmidt rank ($R=3$), regardless of the measurement outcome.

\subsection{Projecting the State onto the MUB Basis}

\noindent The state $\ket{W_{\{0,1,2\}}}$ is an equal-weight superposition of all six ways to distribute the distinct labels $\{0,1,2\}$ among the three particles $A$, $B$, and $C$:

\begin{equation}
\ket{W_{\{0,1,2\}}} = \frac{1}{\sqrt{6}} \sum_{\substack{a,b,c\in\{0,1,2\}\\ \text{all distinct}}} \ket{a}_A\ket{b}_B\ket{c}_C.
\end{equation}

\noindent To measure particle A in an MUB, we must re-express A's computational basis states ($\ket{a}_A$) in terms of the MUB basis vectors ($\ket{e_j}_A$).\\

\noindent Using the inverse relation $\ket{a}_A = \frac{1}{\sqrt{3}}\sum_j \varepsilon_{j,a}\ket{e_j}_A$, we introduce the phase $\varepsilon_{j,a} = \omega^{-(ja+ma^2)}$. Crucially, because $\omega$ is a root of unity, this phase always has a magnitude of 1 ($|\varepsilon_{j,a}| = 1$).\\

\noindent Substituting this back into our superposition gives:

\begin{equation}
\ket{W_{\{0,1,2\}}} = \frac{1}{\sqrt{18}} \sum_{j=0}^{2} \ket{e_j}_A \otimes \ket{\chi_j}_{BC}.
\end{equation}

\noindent Here, $\ket{\chi_j}_{BC}$ represents the unnormalized state of particles B and C if we get outcome $j$ on particle $A$.\\

\noindent Grouping the terms by their basis kets yields:

\begin{equation}
\ket{\chi_j}_{BC} = \varepsilon_{j,0}(\ket{12} + \ket{21}) + \varepsilon_{j,1}(\ket{02} + \ket{20}) + \varepsilon_{j,2}(\ket{01} + \ket{10}).
\end{equation}

\noindent Because all six basis kets are mutually orthogonal and every $\varepsilon$ coefficient has a modulus of 1, calculating the norm of this state is straightforward: $1+1+1+1+1+1 = 6$.

\begin{equation}
\|\chi_j\|^2 = 6.
\end{equation}

\subsection{Measurement Probabilities and the Normalized State}

\noindent Because the norm is independent of $j$, the probability of obtaining any specific measurement outcome $j$ on particle $A$ is perfectly uniform:

\begin{equation}
p(j) = \frac{1}{18}\|\chi_j\|^2 = \frac{6}{18} = \frac{1}{3}.
\end{equation}

\noindent To find the normalized post-measurement state $\ket{\phi_j}_{BC}$, we divide $\ket{\chi_j}_{BC}$ by its norm ($\sqrt{6}$). We can also factor out the global phase $\varepsilon_{j,0}$ to clean up the expression. By defining the relative phases $s_j = \varepsilon_{j,1}/\varepsilon_{j,0}$ and $t_j = \varepsilon_{j,2}/\varepsilon_{j,0}$, the normalized state becomes:

\begin{equation}
\ket{\phi_j}_{BC} = \frac{1}{\sqrt{6}} \Big[ (\ket{12}+\ket{21}) + s_j(\ket{02}+\ket{20}) + t_j(\ket{01}+\ket{10}) \Big].
\end{equation}

\noindent Since all original $\varepsilon$ phases had a magnitude of 1, our new relative phases also sit perfectly on the unit circle ($|s_j| = |t_j| = 1$).

\subsection{The Coefficient Matrix and the Rank-One Trick}

\noindent To find the Schmidt rank of this residual state, we need the eigenvalues of its reduced density matrix, given by $M_j^\dagger M_j$. Reading the coefficients from $\ket{\phi_j}_{BC}$ in the standard basis order $(\ket{0}, \ket{1}, \ket{2})$ yields the coefficient matrix $M_j$:

\begin{equation}
M_j = \frac{1}{\sqrt{6}} \begin{pmatrix} 0 & t_j & s_j \\ t_j & 0 & 1 \\ s_j & 1 & 0 \end{pmatrix}.
\end{equation}

\noindent To find the reduced density matrix, we calculate $M_j^\dagger M_j$. Let us scale the matrix by defining $N = 6 M_j^\dagger M_j$ to remove the fractions.

\begin{align}
N = 6 M_j^\dagger M_j &= 
\begin{pmatrix} 0 & \overline{t_j} & \overline{s_j} \\ \overline{t_j} & 0 & 1 \\ \overline{s_j} & 1 & 0 \end{pmatrix}
\begin{pmatrix} 0 & t_j & s_j \\ t_j & 0 & 1 \\ s_j & 1 & 0 \end{pmatrix} \nonumber \\
&= \begin{pmatrix} 
|t_j|^2 + |s_j|^2 & \overline{s_j} & \overline{t_j} \\ 
s_j & |t_j|^2 + 1 & s_j\overline{t_j} \\ 
t_j & t_j\overline{s_j} & |s_j|^2 + 1 
\end{pmatrix} \nonumber \\
&= \begin{pmatrix} 
2 & \overline{s_j} & \overline{t_j} \\ 
s_j & 2 & s_j\overline{t_j} \\ 
t_j & t_j\overline{s_j} & 2 
\end{pmatrix}.
\end{align}

\noindent Calculating the eigenvalues of this $3 \times 3$ matrix directly would normally result in a complex algebraic mess. However, looking closely at the structure of $N$, we can drastically simplify this using a matrix decomposition trick. \\

\noindent We can separate $N$ into two highly symmetric parts: the standard Identity matrix ($I$) and a \textit{rank-one} matrix formed by the outer product of a column vector $w_j$:

\begin{equation}
N = I + w_jw_j^\dagger \quad \text{where} \quad w_j = \begin{pmatrix} 1 \\ s_j \\ t_j \end{pmatrix}.
\end{equation}

\noindent To verify this, we explicitly calculate the outer product $w_jw_j^\dagger$. Recognizing again that $|s_j|^2 = 1$ and $|t_j|^2 = 1$, we get:

\begin{align}
w_jw_j^\dagger &= \begin{pmatrix} 1 \\ s_j \\ t_j \end{pmatrix} \begin{pmatrix} 1 & \overline{s_j} & \overline{t_j} \end{pmatrix} \nonumber \\
&= \begin{pmatrix} 
1 & \overline{s_j} & \overline{t_j} \\ 
s_j & |s_j|^2 & s_j\overline{t_j} \\ 
t_j & t_j\overline{s_j} & |t_j|^2 
\end{pmatrix} \nonumber \\
&= \begin{pmatrix} 
1 & \overline{s_j} & \overline{t_j} \\ 
s_j & 1 & s_j\overline{t_j} \\ 
t_j & t_j\overline{s_j} & 1 
\end{pmatrix}.
\end{align}

\noindent By adding the $3 \times 3$ Identity matrix $I$, the $1$'s on the main diagonal shift to $2$'s, perfectly recovering our matrix $N$. This separation is the key to the entire proof, as it neatly isolates all the complex relative phases ($s_j$ and $t_j$) inside the $w_jw_j^\dagger$ term.

\subsection{Calculating the Eigenvalues via the Characteristic Equation}

\noindent To find the eigenvalues of the $w_jw_j^\dagger$ matrix, we solve its characteristic equation, $\det(w_jw_j^\dagger - \lambda I) = 0$:

\begin{equation}
\det \begin{pmatrix} 
1-\lambda & \overline{s_j} & \overline{t_j} \\ 
s_j & 1-\lambda & s_j\overline{t_j} \\ 
t_j & t_j\overline{s_j} & 1-\lambda 
\end{pmatrix} = 0.
\end{equation}

\noindent Expanding this determinant along the first row yields:
\begin{align}
(1-\lambda) \left[ (1-\lambda)^2 - (s_j\overline{t_j})(t_j\overline{s_j}) \right] - \overline{s_j} \left[ s_j(1-\lambda) - (s_j\overline{t_j})t_j \right] + \overline{t_j} \left[ s_j(t_j\overline{s_j}) - (1-\lambda)t_j \right] = 0.
\end{align}

\noindent We can significantly simplify the terms inside the brackets by using the fact that our relative phases lie on the unit circle ($|s_j|^2 = 1$ and $|t_j|^2 = 1$):
\begin{align}
(s_j\overline{t_j})(t_j\overline{s_j}) &= |s_j|^2 |t_j|^2 = 1 \nonumber\\
s_j(1-\lambda) - s_j|t_j|^2 &= s_j - s_j\lambda - s_j = -s_j\lambda \nonumber\\
|s_j|^2t_j - (1-\lambda)t_j &= t_j - t_j + \lambda t_j = \lambda t_j.
\end{align}

\noindent Substituting these simplified terms back into the determinant equation:
\begin{align}
(1-\lambda) \left[ (1-\lambda)^2 - 1 \right] - \overline{s_j}(-s_j\lambda) + \overline{t_j}(\lambda t_j) &= 0, \nonumber \\
(1-\lambda)(\lambda^2 - 2\lambda) + \lambda|s_j|^2 + \lambda|t_j|^2 &= 0, \nonumber \\
\lambda(1-\lambda)(\lambda - 2) + \lambda + \lambda &= 0, \nonumber \\
\lambda(-\lambda^2 + 3\lambda - 2) + 2\lambda &= 0, \nonumber \\
-\lambda^3 + 3\lambda^2 - 2\lambda + 2\lambda &= 0. \nonumber \\
\lambda^2(3 - \lambda) &= 0
\end{align}

\noindent The roots of this characteristic equation are exactly $\lambda = 3$ and a degenerate root of $\lambda = 0$. Thus, the eigenvalues of the $w_jw_j^\dagger$ component are $\{3, 0, 0\}$.\\

\noindent Because our full matrix $N$ is defined as $N = I + w_jw_j^\dagger$, adding the Identity matrix simply shifts every eigenvalue up by exactly $1$. This makes the eigenvalues of $N$ equal to $\{4, 1, 1\}$.\\

\noindent Finally, to get back to our actual density matrix ($M_j^\dagger M_j = \frac{1}{6}N$), we divide these eigenvalues by $6$:

\begin{equation}
\lambda_1 = \frac{2}{3}, \quad \lambda_2 = \frac{1}{6}, \quad \lambda_3 = \frac{1}{6}.
\end{equation}

\noindent Because all three eigenvalues are strictly positive, the Schmidt rank of the leftover $B-C$ pair is exactly 3.

\subsection{Result}

\noindent We have shown that for every measurement outcome $j \in \{0,1,2\}$ and every MUB choice $m \in \{0,1,2\}$:

\begin{equation}
\boxed{p(j) = \frac{1}{3}, \quad R=3, \quad \{\lambda_1,\lambda_2,\lambda_3\} = \left\{\frac{2}{3},\frac{1}{6},\frac{1}{6}\right\}}.
\end{equation}


\section{Summary of MUB Measurement Results}
\label{app:mub-summary}

\noindent The following table summarizes the measurement outcomes for all symmetric qutrit $W$-states when particle $A$ is measured in any of the three non-computational Mutually Unbiased Bases (MUBs), indexed by $m \in \{0,1,2\}$. Remarkably, the probabilities, eigenvalues, and resulting Schmidt ranks depend neither on the specific MUB chosen nor on the specific measurement outcome $j \in \{0,1,2\}$.

\begin{table}[H]
\centering
\caption{Summary of measurement outcomes, probabilities, eigenvalues of the reduced density matrix $M^{\dagger}M$, and Schmidt ranks ($R$) for qutrit $W$-states measured in a Mutually Unbiased Basis.}
\renewcommand{\arraystretch}{1.5}
\begin{tabular}{c|c|c|c|c}
\toprule
State Family & Outcome on $A$ & Probability $p(j)$ & Eigenvalues $\{\lambda_i\}$ & $R$ \\
\midrule
Two-same-one-different & \multirow{2}{*}{$\ket{e_j}$} & \multirow{2}{*}{$1/3$} & \multirow{2}{*}{$\left\{\frac{3+\sqrt{5}}{6}, \frac{3-\sqrt{5}}{6}, 0\right\}$} & \multirow{2}{*}{2} \\
($\ket{W_1^{sym}}$ to $\ket{W_6^{sym}}$) & & & & \\
\midrule
All-different & \multirow{2}{*}{$\ket{e_j}$} & \multirow{2}{*}{$1/3$} & \multirow{2}{*}{$\left\{\frac{2}{3}, \frac{1}{6}, \frac{1}{6}\right\}$} & \multirow{2}{*}{3} \\
($\ket{W_{\{0,1,2\}}}$) & & & & \\
\bottomrule
\end{tabular}
\end{table}

\end{document}